\documentclass[aps,preprint]{revtex4}
\usepackage{graphicx,amsmath,amssymb}

\newcommand{\bea}{\begin{eqnarray}}
\newcommand{\eea}{\end{eqnarray}}
\newcommand{\be}{\begin{equation}}
\newcommand{\ee}{\end{equation}}
\newcommand{\lbl}{\label}

\begin{document}

\begin{titlepage}

\title{Qualitatively different collective and single particle\\
dynamics in a supercooled liquid}

\author
{Madhu Priy${\rm a}^{(a)}$, Neeta Bidhoodi, and Shankar P. Das}

\affiliation{School of Physical Sciences,\\
Jawaharlal Nehru University,\\
New Delhi 110067, India.}

\altaffiliation{a: Present address: Institut f\"{u}r Materialphysik
im Weltraum, Deutsches Zentrum f\"{u}r Luft- und Raumfahrt (DLR),
51170 K\"{o}ln, Germany}

\setcounter{equation}{0}

\begin{abstract}

The equations of fluctuating nonlinear hydrodynamics for a two
component mixture are obtained with a proper choice of slow
variables which correspond to the conservation laws in the system.
Using these nonlinear equations we construct the basic equations of
the mode coupling theory (MCT) and consequent ergodic-nonergodic
(ENE) transition in a binary mixture. The model is also analyzed in
the one component limit of the mixture to study the dynamics of a
tagged particle in the sea of identical particles. According to the
existing MCT, dynamics of the single particle correlation is slaved
to that of the collective density fluctuations and hence both
correlations freeze simultaneously at the ENE transition. We show
here from a non-perturbative approach that at the ENE transition,
characterized by the freezing of the long time limit of the dynamic
correlation of collective density fluctuations to a nonzero value,
the tagged particle correlation still decays to zero. Our result
implies that the point at which simulation or experimental data of
self diffusion constant extrapolate to zero would not correspond to
the ENE transition of simple MCT.

\end{abstract}


\maketitle
\end{titlepage}


\section{Introduction}
\label{sec1}

The self-consistent mode coupling theory (MCT) has been a useful
tool for understanding slow dynamics in a dense liquid starting from
the liquid side. The construction of this model involves a basic
feedback mechanism affecting the transport properties of the liquid,
arising from the coupling of slowly decaying density fluctuations in
the supercooled state. The basic result of the model is that as the
density of the liquid increases beyond a critical value, a dynamic
transition from the ergodic liquid state to a nonergodic ideal
glassy state occurs. The long time limit of the time correlation of
density fluctuations is treated as an order parameter for this
transition. This quantity, termed as the nonergodicity parameter
(NEP), makes a discontinuous jump from being zero in the liquid
state to a nonzero positive value at the ergodic-nonergodic (ENE)
transition.

For understanding the mechanism of glass formation in liquids with
simple interaction potentials like hard sphere or Lennard-Jones
type, computer simulation of a  small number of particles moving
under classical laws of motion has been a useful tool
\cite{pastore,woodcock,kob-anderson}. In such simulation studies
often binary mixtures are the system of choice, since they can be
tuned to avoid crystallization of the liquid and thus facilitate the
study of the supercooled state \cite{pastore}. For the two component
systems, the self-consistent MCT with the prediction of an ENE
transition has been studied by several authors \cite{bosse,KB,gtvt}
in the past. The approach adopted there is a straightforward
generalization of the MCT for the one component case. In these works
analysis of experimental and simulation data is made through
schematic models or treating the various non universal parameters in
the theory as freely adjustable for data fitting. The model
equations in these works predict the dynamic transition too
prematurely even when the structural inputs for the MCT was taken
from the simulations. This aspect of the mode coupling model for the
binary fluid is indicated in the computer simulation results
reported in Refs. \cite{nauroth,nauroth-kob} in which the authors
simulated a binary Lennard-Jones system. In this paper we  refer to
these models of binary mixture with the phrase as the ``existing
MCT".

In the current work we present a different formulation of MCT for a
binary mixture by constructing the renormalized perturbation theory
for the dynamics of the collective modes of a two component system.
The collective modes represent the underlying conservation laws of
mass and momentum and we use the equations of  fluctuating nonlinear
hydrodynamics (FNH) for describing their evolution in time. The
basic conservation laws thus play a key role in the construction of
the mode coupling model that we develop here. The self-consistent
MCT is formulated using a Martin-Siggia-Rose (MSR) type field theory
corresponding to the stochastic dynamical equations. Previous works
\cite{PRL84,DM,Das90} using similar techniques for one component
systems have provided important insight in our understanding of MCT
for such systems.  At the simplest level irreversible dynamics of
the slow modes is expressed using bare transport coefficients in the
equations of fluctuating linear hydrodynamics \cite{cohen}. The
transport coefficients represent the role of short time or binary
collision events in producing dissipation. The nonlinear couplings
of the slow modes in the FNH equations represent the role of
correlated motions of the particles at high density and give rise to
renormalization of the bare frictional coefficients. The reversible
part or the Euler terms in the fluctuating equations are obtained
using the Poisson bracket relations between the microscopic
variables. The nonlinearities in the dynamics which give rise to the
feedback mechanism of the MCT and causes the ENE transition are
present in this reversible part, namely the pressure term in the
momentum conservation equation. In the present model formulation for
the binary mixture we focus on the dynamic instability or the ENE
transition as a first step. Within the self-consistent MCT, we
consider the corrections to the transport coefficients using the
mode coupling approximation of dominant density fluctuations and
this is the key ingredient in producing a feedback to the transport
properties. For the binary mixture, couplings to the concentration
fluctuations also become equally relevant. In the MSR theory the
renormalization of both the viscosity and the inter-diffusion are
expressed in a self-consistent form. The possible ENE transition,
allowed by the model equations, is analyzed in terms of the solution
of a set of integral equations for the NEPs. These equations are
obtained from the long time limit of the corresponding correlation
functions constructed from the MSR field theoretic model.

The equal time or structural correlations in a binary fluid are
obtained with a proper free energy functional written in terms of
the slow variables and is a required input for the study of the
dynamics. It is obtained from the coarse graining of the microscopic
Hamiltonian of the binary mixture. Following the method of Langer
and Turski \cite{langer} the momentum density dependent part is
obtained \cite{uh-pre}. The so called interaction part of the free
energy functional is taken in the standard form with expansion in
terms of direct correlation functions \cite{RY}. In the present work
we confine to a strictly Gaussian type free energy functional. The
two point direct correlation functions can be expressed in terms of
static structure factors of the binary mixture through
Ornstein-Zernike relations \cite{frisch}. The thermodynamic
properties of the fluid determined from the interaction potential of
the particles thus enter the formulation of the dynamics. In the
mode coupling model for the dynamics, computation of the mode
coupling integrals appearing in the renormalized transport
coefficients requires the static structure factor of the liquid as
an input. In the one component case the Percus Yevick (PY) structure
factor has been used mostly in similar situations. In case of the
binary mixtures we use the extension of the PY models for a two
component fluid by Lebowitz \cite{lebowitz}. These structure factors
are obtained as a function of the packing fraction $\eta$, size
ratio $\alpha$ ( of species 2 to that of species 1) and the relative
abundance of the species 1 denoted by the variable $x$.

Using the field theoretic formulation for the dynamics of the
collective modes for the binary system, we are able to consider the
one component limit of the mixture by setting the size ratio and
mass ratio of the two species to be unity. The process of
self-diffusion can then be considered by taking the system as a
mixture of a single (tagged) particle with $(N-1)$ particles. In the
existing MCT, the time correlation $\phi(t)$ (say) of collective
density fluctuations couples to time correlation of the single
particle $\phi_s(t)$. Therefore as the $\phi$ freezes at the ENE
transition, so does $\phi_s$ which is simply slaved to the former
and hence the tagged particle diffusion is zero at this point. In
the present work we consider the implications of the ENE transition
on the dynamics of a single particle in a sea of identical
particles.

The paper is organized as follows. In the next section, we define
the proper set of conserved densities for the two component system
and obtain the equations of FNH for the slow variables. In
Sec.~\ref{sec3}, we introduce the MSR field theory  for treating the
nonlinearities in the FNH equations and construction of the
renormalized perturbation theory. Here we demonstrate how the theory
can be renormalized in terms of the elements of  the self-energy
matrix defined with the so called Dyson equation. In Sec.~\ref{sec4}
we discuss the ENE transition and the resulting equations for the
NEPs at the one loop order renormalization. In order to clearly
indicate the difference of the present approach from existing MCT
for binary systems, we also discuss here the approximations involved
in obtaining the latter. In Sec.~\ref{sec5}, we consider the
existing MCT model and discuss the approximations involved in
reaching the same with the use of the MSR approach. In
Sec.~\ref{sec6} we demonstrate through a non perturbative analysis
developed in Ref. \cite{DM09} that the single particle dynamics
decouples from the collective correlation's behavior near the ENE
transition. In the final section we evaluate our results in the
background of the existing MCT for binary systems.

\section{Generalized Hydrodynamics for a binary mixture}
\label{sec2}

The dynamics of the many particle system is studied in terms of the
time evolution of a set of slow modes. The latter arise as a
consequence of underlying conservation laws, broken symmetries
\cite{forster,goldstone}, or due to specific physical property of
the system in consideration. The equations of motion of these
microscopically conserved variables are obtained in terms of
generalized Langevin equations. Using standard formulations
\cite{gene-book3}, the Langevin equations for the coarse grained
densities $\{\psi_i\}$ are obtained in the generalized form
(we adopt the notation that the 
repeated indices are summed over):
\be \label{glgvn} \frac{\partial\psi_\alpha}{\partial t} = \left [
Q_{\alpha\nu}-\Gamma^0_{\alpha\nu} \right ] \frac{\delta
F}{\delta\psi_\nu} +\zeta_\alpha, \ee
where $\zeta_\alpha$ denotes the thermal noise which is assumed to
be Gaussian and white. Correlation of the noise is related through
standard fluctuation dissipation relations (FDRs) to the bare or
short time transport matrix $\Gamma_{\alpha\sigma}^0$ and introduces
the irreversible dynamics for the collective modes.
$\Gamma^0_{\alpha\sigma}$ in Eq. (\ref{glgvn})  is related to the
correlation of the thermal noise $\zeta_{\alpha}$ with FDRs:
\begin{equation}
\label{cnoise-cor} <\zeta_{\alpha}(t)\zeta_{\sigma}(t')> =
2{\beta^{-1}}\Gamma^0_{\alpha\sigma}\delta(t-t').
\end{equation}
where $\beta = (k_{B}T)^{-1}$, is the inverse of temperature $T$
times the Boltzmann constant $k_B$. $F[\psi]$ is identified as the
free energy functional of the local densities $\{\psi_\alpha({\bf
x},t)\}$ and determines the equal time correlations or
susceptibility matrix $\chi_{\alpha\sigma}^{-1}$ with
$\alpha,\sigma\in$ the set of slow modes for the system. $F[\psi]$
is expressed in terms of the slow modes. Thus equilibrium averages
of the fields $\psi_{\alpha}$ at equal times are given by
\begin{equation}
\label{eqt-av} <\psi_{\alpha}\psi_{\sigma}>= \frac{\int
D(\psi)e^{-\beta F[\psi]}\psi_{\alpha}\psi_{\sigma}} {\int
D(\psi)e^{-\beta F[\psi]}},
\end{equation}
where $D(\psi)$ indicates a functional integral over the fields
$\{\psi_{\alpha}\}$. The stationary solution of the Fokker-Planck
equation corresponding to Langevin equations for
fluctuating hydrodynamics is $e^{-\beta F}$. 
In the following we consider the set of equations for a binary
mixture which forms the basis for the model of self-diffusion we
consider here. For the binary mixture elements of the bare transport
matrix include the viscosities and inter-diffusion coefficients.
$Q_{\alpha\nu}=\{\psi_\alpha,\psi_\nu\}$ in Eq. (\ref{glgvn}) is the
Poisson bracket between the slow variables $\psi_\alpha$ and
$\psi_\nu$.

We consider here the FNH equations for a binary mixture of $N_s$
identical particles of species $s$ having mass $m_s$ for $s=1,2$
respectively. $x_s=N_s/N$ is the concentration of the species $s$
and $N=N_{1}+N_{2}$ is the total number of particles. For the binary
system, we consider the following set of collective variables which
are treated as slow due to the underlying microscopic conservation
of the individual mass and sum of the total momentum of the two
species respectively. The individual mass densities $\rho_s$ and the
momentum densities ${\bf g}_s$ for the species $s$ are respectively
defined in terms of microscopic phase space variables as follows
\cite{morozov},
\bea \label{rhos-def}\rho_{s}({\bf x},t) &=&
m_{s}\sum_{i=1}^{N_{s}}\delta({\bf x}-
{\bf R}_{s}^{i}(t)), \\
\label{gs-def} {\bf g}_{s}({\bf x},t) &=& \sum_{i=1}^{N_{s}}{\bf
p}_{s}^{i}\delta({\bf x}
-{\bf R}_{s}^{i}(t)),
\eea
where $m_{1}$ and $m_{2}$ are the masses of particles of species 1
and 2 respectively. The phase space coordinates of position and 
momentum of the $i$-th  particle of the species $s$ are denoted as
$\{R_s^i(t),P_s^i(t)\}$. The individual coarse grained mass
densities, respectively denoted as $\rho_s({\bf x},t)$ are
microscopically conserved. The individual momentum densities ${\bf
g}_1$ and ${\bf g_2}$ are not conserved but total momentum density
defined as,
\be \label{geqns}
 {\bf g}({\bf x},t)={\bf g}_{1}({\bf x},t)+{\bf g}_{2}({\bf x},t)\ee
is conserved. We work here with the following  set of conserved
variables: the total mass density $\rho({\bf x},t)$, total momentum
density ${\rm g}({\bf x},t)$, and the concentration variable $c({\bf
x},t)$ \cite{cohen}. The mass and concentration densities are
defined as follows:
\bea \lbl{totden-def}
\rho({\bf x},t) &=& \rho_{1}({\bf x},t)+\rho_{2}({\bf x},t), \\
\lbl{conc-def}
 c({\bf x},t)&=& x_{2}\rho_{1}({\bf x},t)-x_{1}\rho_{2}({\bf
 x},t).
\eea
We define the fluctuations of $\rho$ and $c$ respectively as
$\delta\rho=\rho-\rho_0$ and $\delta{c}=c$, since the average of $c$
is zero when we consider the mass ratio of the constituent particles
to be unity.

The generalized Langevin equation (\ref{glgvn}), leads to the
equations of motion for the respective coarse grained densities
$\psi_i\equiv\{\rho({\bf x},t),{\bf g}({\bf x},t),c({\bf x},t)\}$
for a binary mixture. Following standard procedures \cite{mybook},
outlined in the Appendix \ref{appendix1} we obtain:
\bea && \label{cont-eqn}
\frac{\partial\rho}{\partial t}+\nabla . {\bf g}=0, \\
&& \label{momt-eqn} \frac{\partial {\rm g}_i}{\partial t}+ {\bf
\nabla}_j \left [ \frac{{\rm g}_i{\rm g}_j}{\rho} \right
]+\rho\nabla_{i} \frac{\delta
F_{U}}{\delta\rho}+c\nabla_{i}\frac{\delta F_{U}}{\delta c} +
L^0_{ij}\frac{{\rm g}_j}{\rho}
=\theta_{i},\\
&& \label{conc-eqn} \frac{\partial c}{\partial t}+{\bf \nabla}\cdot
\left [ c\frac{\bf g}{\rho} \right ] +\gamma_{cc}\nabla^2
\frac{\delta F_U}{\delta c}=f_c. \eea
$F_U$ is the so called potential  part of the free energy functional
$F[\psi]$ introduced in the generalized Langevin equation
(\ref{glgvn}). $F$ is expressed as \be F[\rho,{\bf
g},c]=F_K[\rho,{\bf g}]+F_U[\rho,c], \ee
where the kinetic part $F_K$ (dependent on the current density ${\bf
g}$) is computed from the microscopic Hamiltonian considering the
partition function of the system and following the method of Langer
and Turski \cite{langer,uh-pre} we obtain,
\be \label{fekin} F_{K}[\rho,{\bf g}]=\int{d{\bf x}\frac{{\bf
g}^{2}({\bf x})}{2\rho({\bf x})}}. \ee
$F_U[\rho,c]$ is taken here as a quadratic functional of the fields
$\rho$ and $c$, and is related to the structure of the liquid. This
is expressed in terms of the corresponding direct correlation
functions $\{c_{\rho\rho},c_{\rho{c}},c_{cc}\}$ defined in the
Ornstein-Zernike relations. See Appendix \ref{appendix1} for details
on the structure of the mixture.

The various elements of the bare transport matrix
$\Gamma^0_{\alpha\sigma}$ which appear in the generalized Langevin
equation (\ref{glgvn})  for the binary mixture are defined in
Appendix \ref{appendix1}. Thus $L_{ij}^0$ represents the matrix of
bare or short time viscosities while $\gamma_{cc}$ links to the bare
inter-diffusion coefficient for the mixture. These two dissipative
coefficients are related to the correlation of the Gaussian noises
respectively in Eqs. (\ref{momt-eqn}) and (\ref{conc-eqn}),
\bea \label{cnoise-cor1} \left\langle f_c({\bf x},t)f_c({\bf
x}^{\prime},t^{\prime}) \right\rangle &=&
2{\beta^{-1}}\gamma_{cc}\nabla^2
\delta({\bf x}-{\bf x}^{\prime})\delta(t-t^{\prime}), \\
\label{gnoise-cor} \left\langle \theta_{i}({\bf x},t)\theta_{j}({\bf
x}^{\prime},t^{\prime}) \right\rangle &=& 2{\beta^{-1}}L^0_{ij}
\delta({\bf x}-{\bf x}^{\prime})\delta(t-t^{\prime}),\\
\label{cgnoise-cor} \left\langle f_c({\bf x},t) \theta_i ({\bf
x}^{\prime},t^{\prime}) \right\rangle &=& 0. \eea
Here $\beta^{-1}$ determines the strength of the thermal noise
correlations. For an isotropic system the bare viscosity matrix
$L_{ij}^0$ involves two independent coefficients,
\be \label{visc-tensor} L^0_{ij}= -L_0\nabla_i\nabla_j -\eta_0
(\delta_{ij}{\nabla}^2-\nabla_i\nabla_j),\ee
where $L_0$ and $\eta_0$ respectively denotes the bare or short time
longitudinal and shear viscosities.

Equations (\ref{cont-eqn})-(\ref{conc-eqn}) represent the
dissipative dynamics of the slow modes in a binary mixture due to
nonlinear coupling of the these modes. The mode coupling model for
slow dynamics of a mixture follows from these equations. To focus on
the role of various nonlinearities we note the following terms in
the above set of equations. In Eq. (\ref{momt-eqn}) for the momentum
density ${\bf g}$, the second, third and the fourth term in LHS
represent various contributions from the reversible part of the
dynamics. The second term represents a convective nonlinearity and
ensures Galilean invariance of the equations while the third and the
fourth terms correspond to nonlinear dynamics. Even with a Gaussian
free energy functional defined in Eq. (\ref{fepot}) these two terms
give rise to a nonlinear coupling of $\rho$ and $c$. In Eq.
(\ref{conc-eqn}) for concentration $c$, the second term in the LHS
represents reversible dynamics and is the only nonlinearity. The
third term represents the dissipative part that corresponds to a
diffusive mode. The only nonzero Poisson bracket of the
concentration variable is $\{c,{\rm g}_i\}$ and since the the
functional derivative of $F$ with ${\rm g}_i$ is ${\rm g}_{i}/\rho$,
the only possible coupling in the reversible part of the $c$
equation is between $c$ and ${\rm g}_i/\rho$. This is an important
point to note and will be useful when we consider the
renormalization of the dynamics due to the nonlinearities. The role
of $1/\rho$ nonlinearities in the FNH equations will be ignored in
this work to primarily focus on the ENE transition in particular.

At the linear level the Eqs. (\ref{cont-eqn})-(\ref{conc-eqn}) of
fluctuating hydrodynamics for the mixture involve the characteristic
speeds $c_0$ and $\upsilon_0$ which are respectively expressed in
terms of equilibrium structure factors,
$c_0^2=\rho_0\chi_{\rho\rho}^{-1}$ and
$\upsilon_0^2=\rho_0\chi_{\rho{c}}^{-1}$. The dissipative equations
of linearized dynamics also include the bare transport coefficients
which are respectively the longitudinal viscosity $L_0$, the
inter-diffusion coefficient, $\nu_0=\gamma_{cc}\chi_{cc}^{-1}$. The
effects of the nonlinearities in the above FNH equations are
accounted through renormalization of the bare transport coefficients
($L_0$, $\nu_0$) as well as the speeds ($c_0$, $\upsilon_0$). In
particular, corrections  of $L_0$ and $\nu_0$ due to the slowly
decaying hydrodynamic modes give rise to a nonlinear feedback
mechanism which is key to producing the slow dynamics of the MCT.


\section{Martin-Siggia-Rose field theory}
\label{sec3}

In this section we describe the computation of the correlation and
response functions of the conserved slow modes which are the prime
quantities in describing the dynamics of the binary mixture and
possible ENE transition in the system. The consequences of the
nonlinearities in the equations of motion, {\em i.e.}, the
generalized Langevin equations for the slow variables are worked out
by using graphical methods of field theory. In the present work the
renormalized perturbation theory is developed in self-consistent
form which is particularly useful for the discussion of the mode
coupling model and the consequent slow dynamics. We follow here
closely the methodology developed in Ref. \cite{DM} using the
standard approach of MSR field theory
\cite{msr1,msr2,msr3,msr4,hca}. We describe the scheme briefly below
and for more technical details we refer the reader to Ref.
\cite{DM}. The renormalized theory for the binary mixture dynamics
is developed in terms of the correlation functions and response
functions respectively given by,
\bea \label{g-function} G_{\alpha\beta}(12)=
\langle\psi_{\alpha}(1)\psi_{\beta} (2)\rangle, \\
\label{r-function} G_{\alpha\hat{\beta}}(12)= \langle
\psi_{\alpha}(1) \hat{\psi}_{\beta}(2)\rangle. \eea
The averages denoted here by the angular brackets are functional
integrals over all the fields weighted by $e^{-{\cal A}}$. The
action ${\cal A}$  is a functional of the field variables
$\{\psi_i\}$ and the corresponding conjugate hatted fields
$\{\hat{\psi}_i\}$ introduced in the MSR filed theory. Using the
equations of motion (\ref{cont-eqn})-(\ref{conc-eqn}) for the set of
slow modes  $\{\rho,c,{\bf g}\}$, the MSR action functional is
obtained in the Appendix \ref{appendix2} as given in Eq.
(\ref{Aaction-MSR}).
The correlation and response functions in the MSR field theory,
respectively given by Eqs. (\ref{g-function}) and
(\ref{r-function}), are suitably organized in terms of their
contributions from the Gaussian and non-Gaussian parts of the action
functional ${\cal A}$. Using the polynomial expansions of the linear
and nonlinear kernel terms in the equations of motion the action
functional ${\cal A}$ in Eq. (\ref{Aaction-MSR}) is put in a
schematic form
\begin{eqnarray}
\label{msr-aschem} {\cal A}_{U}[\Psi,\hat{\Psi}]
&=&\frac{1}{2}\sum_{1,2}\Psi(1)G_{0}^{-1}(12)\Psi(2)+\frac{1}{3}\sum_{1,2,3}
V(123)\Psi(1)\Psi(2)\Psi(3)\nonumber\\
&+&\frac{1}{4}\sum_{1,2,3}
V(1234)\Psi(1)\Psi(2)\Psi(3)\Psi(4)-\sum_{1}\Psi(1)U(1).
\end{eqnarray}
In the above expression the set of slow modes $\{ \psi_\alpha \}$
are represented in terms of a vector field $\Psi(1)$ having the
different fields as its components. The nonlinearities in the
equations of motion (\ref{cont-eqn})-(\ref{conc-eqn}) give rise to
non-Gaussian terms in the action Eq. (\ref{msr-aschem}) involving
products of three or more field variables. The corresponding vertex
functions $V(123)$, etc., (see Eq. (\ref{msr-aschem}) for the MSR
action)  are defined to be symmetric under the exchange of the
indices. The simplest level form of the correlation functions are
zeroth order quantities 
denoted by $G_0$ corresponding to the the
action which is only quadratic order in the fields, all higher order
vertices being ignored. Keeping only the Gaussian terms in the
action functional (\ref{Aaction-MSR}), the matrix $G^{-1}_0$ defined
in Eq. (\ref{msr-aschem}) is obtained in the block form
\begin{equation}
\label{g0-inv} G_0^{-1} = \left [
\begin{array}{cc}
\bigcirc  & {\cal B}_0^{\dag} \\
{\cal B}_0 & {\cal C}_0\\
\end{array}
\right ]
\end{equation}
where the elements of matrix ${\cal B}_0$ are provided in Table
\ref{table1} and the matrix ${\cal B}_{0}^\dag$ is the transpose and
complex conjugate of the matrix ${\cal B}_{0}$. The matrix ${\cal
C}_0$ is defined as,
\begin{equation}\label{C0-matrix}
{[{\cal C}_0]}_{\hat{\mu}\hat{\nu}}=
2\beta^{-1}\delta_{\hat{\mu}\hat{\nu}} \left [
\delta_{\hat{\mu},\hat{\rm g}}L_{0} +
\delta_{\hat{\mu},\hat{c}}\gamma_{cc} \right ].
\end{equation}
The $\bigcirc$ in RHS of Eq. (\ref{msr-aschem}) represents the null
matrix with all its elements equal to zero. The role of the
non-Gaussian terms is to renormalize the correlation functions of
the Gaussian theory and is expressed in a perturbation series
expansion in terms of the corresponding vertices. The diagrammatic
methods of field theories are used for this purpose. In the
following we use the fluctuation dissipation relations
(\ref{fdt-q1})-(\ref{fdt-q3}) to obtain important conclusions on the
renormalized theory from a nonperturbative approach.

\subsection{Fluctuation-dissipation Relations}

We now demonstrate that the correlation and response functions
defined in Eqs. (\ref{g-function}) and (\ref{r-function}) are
related through a set of fluctuation-dissipation relations. The
derivation of these FDRs are based on the symmetry of the MSR action
under time reversal transformation \cite{ABL}. In Appendix
\ref{appendix1} we demonstrate that the MSR action
(\ref{Aaction-MSR}) remains invariant under the following time
transformation rules of the field $\psi_i$ and its hatted conjugate
$\hat{\psi}_i$
\begin{eqnarray}
\psi_{i}({\bf x},-t) &\rightarrow& \epsilon_{i}\psi_{i}({\bf
x},t), \nonumber\\
\label{ttr-r1} \hat{\psi}_{i}({\bf x},-t)&\rightarrow&
-\epsilon_{i}\left[\hat{\psi}_{i}({\bf x},t)-i\beta\frac{\delta
F}{\delta \psi_{i}({\bf x},t)}\right].
\end{eqnarray}
Applying this symmetry corresponding to the field $\psi_i\equiv{\rm
g}_i$, we obtain
\begin{eqnarray}
{\rm g}_{i}({\bf x},-t) &\rightarrow&  -{\rm g}_{i}({\bf x},t),\nonumber\\
\label{ttr-gt} \hat{\rm g}_{i}({\bf x},-t) &\rightarrow& \hat{\rm
g}_{i}({\bf x},t)-i\beta{{\mathrm{v}}_i({\bf x},t)}.
\end{eqnarray}
We denote the functional derivative of the free energy functional
$F$ with the field $\psi({\bf x},t)$ as
\be \zeta_\psi({\bf x})=\frac{\delta F}{\delta \psi({\bf x})}~, \ee
so that  $\zeta_{{\rm g}_i}=({\delta F}/{\delta {\rm g}_i({\bf
x},t)}) = {\rm g}_{i}({\bf x},t)/\rho({\bf x},t) =
{\mathrm{v}}_i({\bf x},t)$. Applying the above transformation rules
to the correlation of $\hat{\rm g}_i({\bf x}_1,-t_1)$ with a field
$\varphi({\bf x}_{2},t_{2})$ we obtain
\begin{equation}
\left\langle\hat{\rm g}_i({\bf x}_1,-t_1)\varphi({\bf x}_2,t_2)\right\rangle=
\left\langle\hat{\rm g}_i({\bf x}_1,t_1)\varphi({\bf x}_2,t_2)\right\rangle-
i\beta\left\langle
{\mathrm{v}_i}({\bf x}_1,t_1)\varphi({\bf x}_2,t_2)\right\rangle.
\end{equation}
For $t_{1}>t_{2}$, the LHS is zero due to causality
and obtains
\begin{equation}
\label{fdt-r1} G_{{\mathrm{v}_i}\varphi}({\bf x},t)=
-i\beta^{-1}G_{\hat{\rm g}_i\varphi}({\bf x},t),
\end{equation}
where ${\bf x}={\bf x}_{1}-{\bf x}_{2}$ and $t=t_{1}-t_{2}$. Since
the response functions by definition are time retarded due to
causality principle, for the spatially Fourier transformed
correlation function from the corresponding response function we
obtain
\begin{equation}
\label{fdt-q1}G_{{\mathrm{v}_i}\varphi}({\bf q},\omega)=
-2\beta^{-1}{\rm Im} G_{\hat{\rm g}_{i}\varphi}({\bf q},\omega).
\end{equation}
The ${\bf v}$ field has been introduced in the formulation to deal
with the $1/\rho$ nonlinearity in the equations of motion. In the
case of a one component liquid the latter plays a crucial role
\cite{DM} in cutting off the sharp ergodic-nonergodic (ENE)
transition in which  density correlation function freezes at a
nonzero value in the long time limit. In the present work we ignore
the ergodicity mechanisms to focus on the implications of the ENE
transition, in particular, in the binary system. Thus, we ignore the
role of the $1/\rho$ nonlinearity and treat ${\bf g}$ and ${\bf v}$
with the linear relation ${\bf g}=\rho_0 {\bf v}$ and work with the
set of fields $\{{\bf g},\rho,c\}$. With this approximation the FDR
given by Eq. (\ref{fdt-q1}) reduces to
\begin{equation}
\label{fdt-q1s} G_{{\rm g}_{i}\varphi}({\bf q},\omega)=
-2\beta^{-1}\rho_{0}{\rm Im} G_{\hat{\rm g}_{i}\varphi}({\bf
q},\omega)~.
\end{equation}
Applying the same symmetries respectively for $\psi_i = \rho$ and
$c$ we obtain the following set of FDRs:
\bea \label{fdt-q2} G_{\zeta_c{\varphi}}({\bf q},\omega) &=&
-2\beta^{-1}{\rm Im} G_{\hat{c}\varphi}({\bf q},\omega), \\
\label{fdt-q3} G_{\zeta_\rho{\varphi}}({\bf q},\omega) &=&
-2\beta^{-1}{\rm Im} G_{\hat{\rho}\varphi}({\bf q},\omega). \eea

\subsection{Renormalization}

The role of the non-Gaussian parts of the action ${\cal A}$ on the
correlation functions are quantified in terms of the self-energy
matrix $\Sigma$ which shows up in the equation satisfied by the
response functions and that satisfied by the correlation functions.
The self-energy matrix ${\bf \Sigma}$ is defined through the
Schwinger-Dyson equation
\begin{equation}
\label{full-Deqn}
{\bf G}^{-1} = {\bf G_0}^{-1} - {\bf \Sigma},
\end{equation}
where ${\bf G_0}$ represents the Gaussian counterpart of ${\bf G}$
obtained by keeping only up to quadratic terms in the action ${\cal
A}$. The matrix equations represented  by Eq. (\ref{full-Deqn}) are
solved to obtain the corresponding correlation and response
functions in MSR field theory. For example, from the set of
equations denoted by (\ref{full-Deqn}) we obtain that the response
functions satisfy:
\begin{equation} \label{dyson-resp}
\left[(G_{0}^{-1})_{\hat{\alpha}\mu}(13)
-\Sigma_{\hat{\alpha}\mu}(13) \right]G_{\mu\hat{\beta}}(32)= \delta
(12)\delta_{\hat{\alpha}\hat{\beta}}.
\end{equation}
Using diagrammatic methods the self-energies
$\Sigma_{\hat{\alpha}\mu}$ as well as
$\Sigma_{\hat{\alpha}\hat{\mu}}$ are expressed in perturbation
theory in terms of the two-point correlation and response functions.
The block form of the inverse Green's function matrix, both in the
zeroth order and in the fully nonlinear theory, have a symmetric
structure in the parts representing the response functions. Since
the fields $\psi$ and $\hat{\psi}$ are real, it readily follows from
the MSR action functional that
\be {\cal A}^*[\psi,\hat{\psi}]={\cal A}[\psi,-\hat{\psi}], \ee
and it is straightforward to show \cite{DM} that the response
function $G_{\hat{\alpha}\beta}$ satisfy the relation,
\begin{equation}
\label{resp-fsym} G_{\hat{\alpha}\beta}({\bf
q},\omega)=-G_{\beta\hat{\alpha}}^{*}({\bf q},\omega).
\end{equation}
From the matrix form (\ref{g0-inv}) for the $G_0^{-1}$ matrix it is
obvious that this is satisfied at the zeroth order. From the Dyson
equation (\ref{full-Deqn}) it therefore also follows that the
self-energies satisfy the relation
\begin{equation}
\label{self-fsym} {\Sigma}_{\hat{\alpha}\beta}({\bf
q},\omega)=-{\Sigma}_{\beta\hat{\alpha}}^{*}({\bf q},\omega).
\end{equation}
Analyzing the structure of the full Green's function matrix from the
Dyson equation in the Appendix \ref{appendix3} the response part of
$G_{\alpha\hat{\beta}}$ is obtained as
\be \label{resp-matx} G_{\alpha\hat{\beta}}=
\frac{N_{\alpha\hat{\beta}}}{{\cal D}}. \ee
The various elements of matrix $N_{{\alpha}\hat{\beta}}$ for the
case of a binary mixture are obtained in Appendix \ref{appendix3}.
The denominator ${\cal D}$ in the RHS of Eq. (\ref{resp-matx}) is
obtained in Eq. (\ref{denom-resp}). The $N_{\hat{\alpha}\beta}$
matrix satisfies the relation $N_{\hat{\alpha}\beta}({\bf
q},\omega)=N_{\beta\hat{\alpha}}^{*}({\bf q},\omega)$. The various
renormalized transport coefficients  appearing in the RHS of Eq.
(\ref{resp-matx}) are expressed in terms of the corresponding
response self-energies as listed in Eqs.
(\ref{renorm-visc})-(\ref{renorm-nup}). The correlation functions of
the physical, un-hatted field variables are defined as,
\begin{equation}
\label{eq:84} G_{\alpha\beta}=-\sum_{\mu\nu }
G_{\alpha\hat{\mu}}{\cal C}_{\hat{\mu}\hat{\nu}}
G_{\hat{\nu}\beta}~,
\end{equation}
where Greek letter subscripts take values $\rho ,c$ and the
longitudinal components of the vector field  ${\bf g}$. The
self-energy matrix ${\cal C}_{\hat{\mu}\hat{\nu}}$ is listed in
Table \ref{table2}. The double-hatted self-energies
$\Sigma_{\hat{\mu}\hat{\nu}}$ as well as ${\cal
C}_{\hat{\mu}\hat{\nu}}$ both vanish if either index corresponds to
$\hat{\rho}$, since there is no noise or nonlinearity in the
continuity equation (\ref{cont-eqn}). Therefore from the general
structure of Eq. (\ref{eq:84}) of the correlation and that of Eq.
(\ref{resp-matx}) for response functions we obtain
\begin{equation}
\label{eq:mixed} G_{\alpha\beta}=\frac{1}{{\cal D}{\cal D}^*}
\sum_{\mu\nu} N_{\alpha\hat{\mu}}{\cal C}_{\hat{\mu}\hat{\nu}} N_{\hat{\nu}\beta}~.
\end{equation}
From the above expression it is clear that the renormalized
correlation functions involve both response type self-energies
($\Sigma_{\psi\hat{\psi}}$) as well as correlation type
self-energies ($\Sigma_{\hat{\psi}\hat{\psi}}$), the latter being
present in the matrix ${\cal C}_{\hat{\psi}\hat{\psi}}$. In order to
demonstrate that the renormalized correlation functions can be
expressed in terms of a set of renormalized transport coefficients
we therefore need to establish a relation between the corresponding
set of response and correlation type self-energies renormalizing the
same transport coefficient. Here the fluctuation dissipation
relations (\ref{fdt-q1s})-(\ref{fdt-q3}) between correlation and the
response functions of the MSR field theory prove very useful. We are
able to do this at the non-perturbative level here but in the
hydrodynamic limit of small wave-vectors and frequencies.

\subsection{Nonperturbative Analysis}
\label{sec-nonpert}

We consider the FDRs (\ref{fdt-q1s})-(\ref{fdt-q3}) obtained in the
previous section to link the correlation and response self-energies.
Using the Eqs. (\ref{resp-fsym}) and (\ref{eq:84}) respectively for
the response function $G_{\hat{g}_{i}\varphi}$ and correlation
function $G_{g_{i}\varphi}$  in the FDRs (\ref{fdt-q1s}) and
(\ref{fdt-q2}) we obtain the following set of relations:
\bea \label{sreln-1} && N_{{\rm g}_{i}\hat{\alpha}}{\cal
C}_{\hat{\alpha}\hat{\gamma}} =-i\beta^{-1}\rho_{0}({\cal D}
\delta_{\hat{\rm g}_{i}\hat{\gamma}}+N^*_{\hat{\rm
g}_{i}\varphi}G^{-1}_{\varphi\hat{\gamma}}), \\
\label{sreln-2} && \{\chi_{\rho
c}^{-1}N_{\rho\hat{\alpha}}+\chi_{cc}^{-1} N_{c\hat{\alpha}} \}
{\cal C}_{\hat{\alpha}\hat{\gamma}} = -i\beta^{-1}( {\cal
D}\delta_{\hat{c}\hat{\gamma}} +
N^*_{\hat{c}\varphi}G_{\varphi\hat{\gamma}}^{-1}). \eea
The results in Eqs. (\ref{sreln-1}) and (\ref{sreln-2}) are further
analyzed in the Appendix \ref{appendix2} to obtain a set of
relations between the response and correlation self-energies. In the
hydrodynamic limit which corresponds to the small wave numbers ($q$)
and small frequencies ($\omega$) we obtain,
\begin{eqnarray}
\label{bvr}
\gamma_{\hat{\rm g}\hat{\rm g}} &=&
2\beta^{-1}\rho_{0}\gamma_{\hat{\rm g}{\rm g}}^{\prime}~,\\
\label{dcr}
\gamma_{\hat{c}\hat{c}} &=& 2\beta^{-1}
\frac{\gamma_{\hat{c}\rho}^{\prime}}{\chi_{\rho c}^{-1}}
=2\beta^{-1}\frac{\gamma_{\hat{c}c}^{\prime}}{\chi_{c c}^{-1}}~.
\end{eqnarray}
The quantities
$\{\gamma_{\hat{\alpha}\beta},\gamma_{\hat{\alpha}\hat{\beta}}\}$,
which appear in the above relations are coming from the leading
order contributions to the corresponding self-energies
$\{\Sigma_{\hat{\alpha}\beta},\Sigma_{\hat{\alpha}\hat{\beta}}\}$.
These leading order wave-vector dependence for the different
self-energies are listed in Appendix \ref{appendix3}. Justifications
for using the properties of the different vertex functions $V$'s in
the action $(\ref{msr-aschem})$ are given in the Appendix
\ref{appendix2}. For the off-diagonal elements of the ${\cal
C}_{\hat{\alpha}\hat{\delta}}$ matrix we also obtain,
\be
 \label{ghgh3} \gamma_{\hat{c}\hat{\rm g}} =
2\beta^{-1}\frac{\gamma_{\hat{\rm g}c}^{\prime}} {\chi_{cc}^{-1}}=
2\beta^{-1}\frac{\gamma_{\hat{\rm g}\rho}^{\prime}} {\chi_{\rho
c}^{-1}}~. \ee
The above relations between the self-energies are important for the
renormalizability of the theory in terms of redefined transport
coefficients. The renormalized longitudinal viscosity coefficient
$L(q,z)$ is obtained in the form
\bea L(q,z) &=& L_0(q) +
i\gamma_{\hat{\rm g}{\rm g}}^{\prime}({\bf q},z)~,\nonumber\\
\label{renp-visc} &=& L_0(q) + {\beta}{\rho^{-1}_{0}} \int_0^\infty
dt e^{izt} \gamma_{\hat{\rm g}\hat{\rm g}}({\bf q},t)~. \eea
From Eq. (\ref{dcr}), we see that in the hydrodynamic limit both
$\gamma_{\hat{c}\rho}^{\prime}$ and $\gamma_{\hat{c}c}^{\prime}$ are
related to the self-energy $\gamma_{\hat{c}\hat{c}}$. This has
important implication in the renormalization of the two transport
coefficients appearing in the correlation function matrix. From Eqs.
(\ref{renorm-nu}) and (\ref{renorm-nup}) and using the definitions
of $\gamma_{\hat{c}c}$ and $\gamma_{\hat{c}\rho}$, we obtain
respectively the renormalized expressions for $\nu(q,\omega)$ and
$\nu'(q,\omega)$ as
\begin{eqnarray}
\label{renp-mu} \nu(q,\omega)&=&
\chi_{cc}^{-1}\gamma_{0}(q)+\gamma_{\hat{c}c}^{\prime}(q,\omega)~,\\
\label{renp-mup} \nu'(q,\omega)&=&
 \chi_{{\rho}c}^{-1}\gamma_{0}(q)+\gamma_{\hat{c}\rho}^{\prime}(q,\omega)~.
\end{eqnarray}
We denote $\gamma_{cc}$ as $\gamma_0$. The renormalized quantities
$\nu$ and $\nu'$ are respectively expressed as
$\chi_{cc}^{-1}\gamma$ and $\chi_{{\rho}c}^{-1}\gamma$ in terms of a
single renormalized quantity whose Laplace transform is defined as
\be
\label{renp-gam0}
\gamma({\bf q},z) =
\gamma_{0}({\bf q})+{\beta}\int_0^\infty dt e^{izt}
\gamma_{\hat{c}\hat{c}}({\bf q},t)~,
\ee
involving the self-energy $\Sigma_{\hat{c}\hat{c}}$. To summarize,
in hydrodynamic limit, the correlation and the response functions in
the fully nonlinear theory are obtained in terms of the renormalized
transport coefficients. The renormalization of the thermodynamic
quantities $c_{0}^{2}$ and $\upsilon_{0}^{2}$  follows from Eqs.
(\ref{renorm-s1})-(\ref{renorm-s2}). In Appendix \ref{appendix2} we
demonstrate that for the Gaussian free energy considered in the
present work these corrections are higher orders in $q$ and vanish
in the hydrodynamic limit.

\section{The Ergodic-Nonergodic transition}
\label{sec4}

We have identified above the respective self-energy matrix elements
which contribute to the renormalized transport coefficients.  The
correlation and response functions of the fully nonlinear theory are
expressed in terms of these renormalized quantities. The next step,
in this is to express  these self-energies in terms of the
correlation functions. This gives rise to a self-consistent scheme
in which correlation functions  satisfy nonlinear equations
involving memory functions which are expressed in terms of the
correlation functions themselves. This essentially constitutes the
feedback mechanism of MCT and has been used extensively  for
understanding the slow dynamics in a dense supercooled liquid. As
the density or the packing fraction of the system is increased,
there is a critical density at which the density correlation
function does not decay to zero in the long time limit and this is
defined as an ENE transition. In the present case of the binary
mixture, the transition is characterized by freezing of the
correlations of
$\rho$ and $c$. The ENE transition in the mixture is characterized
by correlation functions having a nonergodic solution, {\em i.e.},
the long time limit of the different elements of the matrix of
correlation functions $G_{\sigma\mu}(q,t)$, where $\sigma,\mu
\in\{\rho,c\} $, are nonzero for all wave number $q$. This is
equivalent to having the corresponding Laplace transform
$G_{\sigma\mu}(q,z) \sim f_{\sigma\mu}(q)/z$ developing a pole at
$z=0$. Equivalently, the Fourier transform behaves like
$G_{\sigma\mu}(q,\omega) \sim 2\pi\delta(\omega){f_{\sigma\mu}(q)}$.
In the following, the liquid is considered to have an ENE transition
when a set of nonzero values are obtained for the
$f_{\sigma\mu}(q)$'s which are therefore termed as the nonergodicity
parameters.

\begin{figure}[h]
\includegraphics[height=8.0cm]{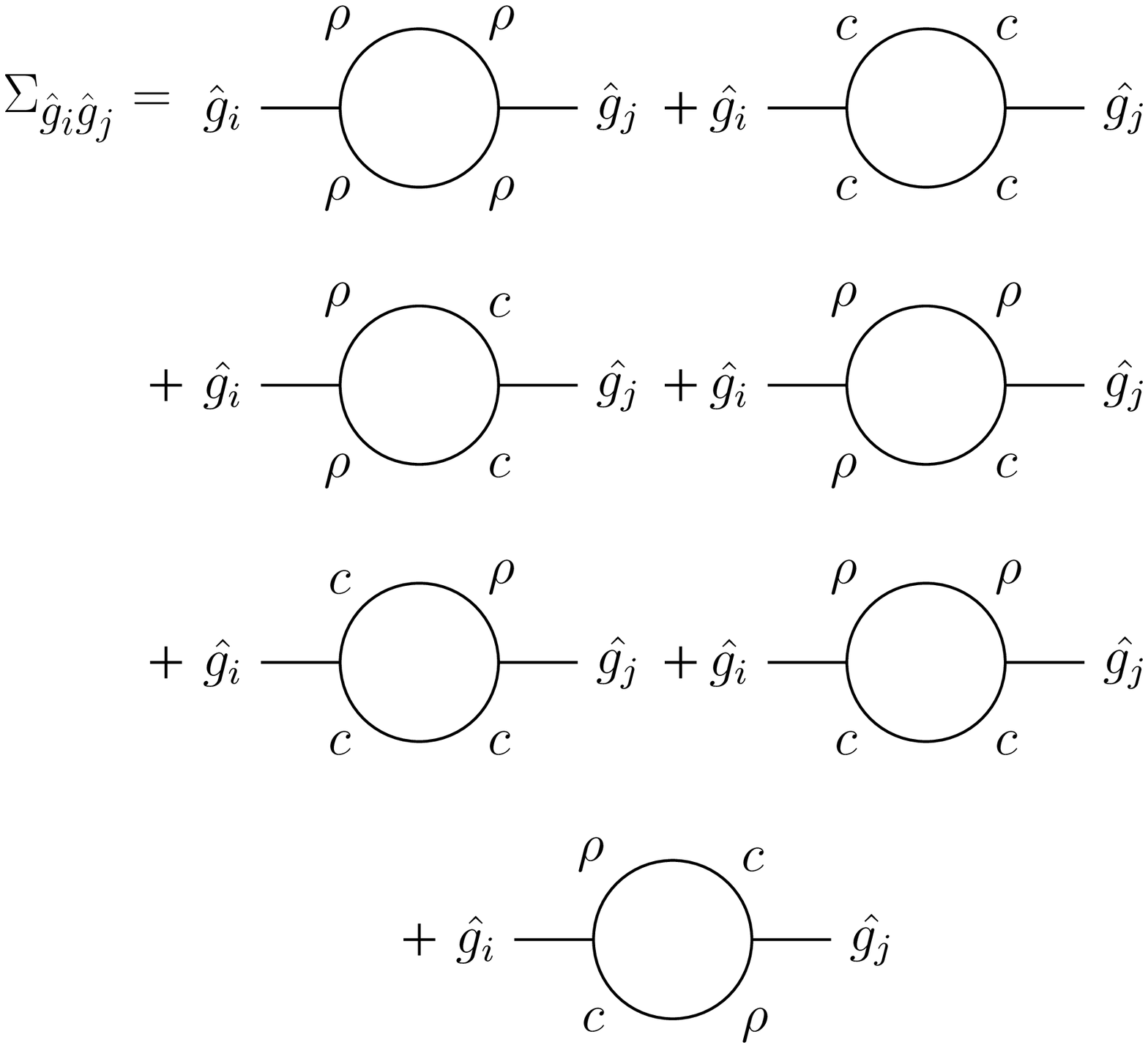}
\caption{One loop diagrams for $\Sigma_{\hat{g}_{i}\hat{g}_{j}}$
with vertices involving nonlinear couplings of density
fluctuations.} \label{fig1}
\end{figure}
\begin{figure}[h]
\includegraphics[height=2.0cm]{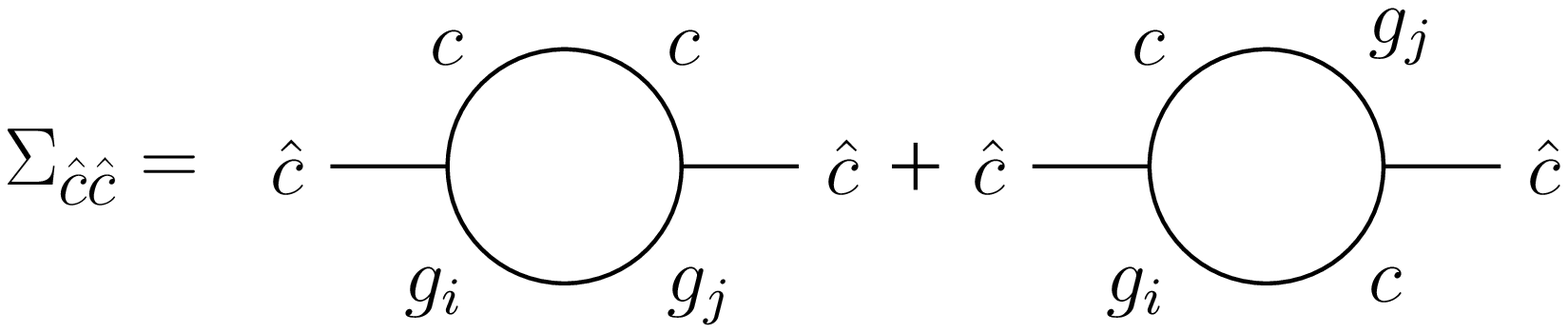}
\caption{One loop contributions to $\Sigma_{\hat{c}\hat{c}}$.}
\label{fig2}
\end{figure}
\begin{figure}[h]
\includegraphics[height=15cm]{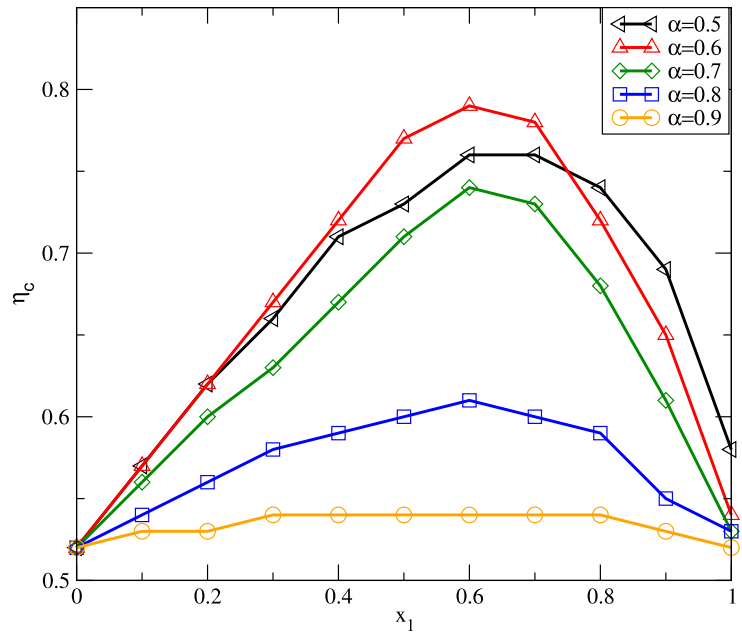}
\caption{The critical packing fraction $\eta_c$ (see text for
definition) for the binary mixture as obtained from the solution of
Eq. (\ref{nep-eqns}) vs. concentration $x_1$ of smaller sized
particles for different size ratios $\alpha$, defined here as the
ratio of smaller to bigger sized particles (having same mass) of the
mixture.} \label{fig3}
\end{figure}

Using the FDRs given by Eqs. (\ref{fdt-q2}) and (\ref{fdt-q3}), the
expressions for the Fourier transform of the correlation functions
are obtained in terms of the Laplace transforms of the correlation
functions,
\be \label{ltc}G_{\alpha\varphi}(q,z)=\frac{1}{\cal D}
\sum_{\nu}N_{\alpha\hat{\nu}}(q,z)\chi_{\nu\varphi}~. \ee
The repeated index $\nu$ in Eq. (\ref{ltc}) is summed over the set
$\{\rho,c\}$. The above equation is rearranged to a form
\be \label{td1}{\cal D} \sum_{\alpha}N^{-1}_{\hat{\mu}\alpha}(q,z)
G_{\alpha\varphi}(q,z)=\chi_{\mu\varphi}(q,z)~. \ee
To sort out the dependence on structure and the dynamics we define
normalized correlation functions $\phi_{\alpha\sigma}$ as,
\be \label{defn-psissp} \phi_{\alpha\sigma}(q,t) =
\frac{G_{\alpha\sigma}(q,t)}{\sqrt{\chi_{\alpha\alpha}\chi_{\sigma\sigma}}}~.
\ee
The corresponding nonergodicity parameter (NEP) is then defined as
the long time limit of the correlation functions
$\phi_{\alpha\sigma}$, {\em i.e.}, the NEPs $f_{\alpha\sigma}$ are
obtained as
\be f_{\alpha\sigma}(q)=\lim_{t\rightarrow\infty}
\phi_{\alpha\sigma}(q,t)=\lim_{z\rightarrow
0}z\phi_{\alpha\sigma}(q,z)~. \ee
Taking the long time limit of the equations given by (\ref{ltc}), we
get the following equations for the NEPs
\be \label{nep-eqns} \frac{f_{\alpha\sigma}(q)}{1-f_{\alpha
\sigma}(q)} =\frac{a_{\alpha\sigma}L(q)}
{\tilde{c_0}^2+(1-a_{\alpha\sigma})L(q)}~. \ee
The $a_{\alpha\sigma}$'s are obtained in terms of static
correlations: $a_{\rho\rho}=1$, $a_{\rho{c}}=\chi$, and
$a_{cc}=\chi^2$ where
$\chi(k)=\chi_{\rho{c}}(k)/\sqrt{\chi_{\rho\rho}(k)\chi_{cc}(k)}$.
$a_{\alpha \sigma}$ is symmetric in $\rho$ and $c$. The quantity
$\tilde{c}_0$ is obtained in terms of the sound speed $c_0$ as,
\be \tilde{c}_0^2=c_0^2+v_0^2(\chi_{\rho c}/\chi_{\rho\rho}).\ee
The function $L(q)$ in the right hand side of Eq. (\ref{nep-eqns})
is the long time limit of the renormalized memory function $L(q,z)$
of the generalized viscosity. The memory function $L(q,z)$ involves
a factor $q^2$ following from conservation laws. The memory
functions for $L(q,t)$ as a nonlinear functional of the
$f_{\alpha\sigma}$'s. In the standard mode coupling theory approach
\cite{DM,beng,leth,rmp} the integral Eqs. (\ref{nep-eqns}) are
closed by treating $L(q)$ as a nonlinear functional of the
$f_{\alpha\sigma}$. This is determined by taking the long time limit
of the corresponding self energies $\Sigma_{\hat{g}\hat{g}}$. Using
the vertex functions appearing in the MSR action functional
(\ref{Aaction-MSR}) we show in the Appendix \ref{appendix3} that the
relevant diagrams (shown in Fig. \ref{fig1}) involving the slowly
decaying correlations of $\rho$ and $c$ obtain the following one
loop contribution for $L(q,t)$
\be \label{oneloop-L} L(q,t)=\frac{\beta^{-1}}{2\rho_{0}q^2}
\int\frac{d{\bf k}}{(2\pi)^{3}} V_{\hat{\rm g}\sigma\sigma^{\prime}}
({\bf q},{\bf k},{\bf k_{1}}) V^{*}_{\hat{\rm
g}\mu\mu^{\prime}}({\bf q},{\bf k},{\bf k_{1}})
\varepsilon_{\sigma\mu}(k)\varepsilon_{\sigma^{\prime}\mu^{\prime}}(k_1)
G_{\sigma\mu}(k,t) G_{\sigma^{\prime}\mu^{\prime}}(k_{1},t)~, \ee
where the repeated indices $\sigma,\mu,\sigma^{\prime}$, and
$\mu^{\prime}$ are summed over the set $\{\rho,c\}$. The vertex
function $V_{\hat{g}\sigma\mu}$ is given by
\be V_{\hat{g}\sigma\mu}({\bf q},{\bf k},{\bf k_{1}})=[\hat{\bf
q}.{\bf k} c_{\sigma\mu}(k)+\hat{\bf q}.{\bf k}_{1} c_{\sigma\mu}
(k_{1})]~, \ee
and ${\bf k}_{1}={\bf q}-{\bf k}$. $\varepsilon_{\sigma\mu}(k)$ is
respectively equal to 1 or $\chi(k)$ for $\sigma=\mu$ and otherwise.
The corresponding renormalization to the diffusion memory kernel
$\gamma(q,t)$ is obtained from the self energy element
$\Sigma_{\hat{c}\hat{c}}$. The relevant one loop diagrams (shown in
Fig. \ref{fig2}) involving the slowly decaying correlations of
$\rho$ and $c$ obtain the following one loop contribution
\be \label{oneloop-D} \gamma(q,t)=\frac{2}{\rho_0^2}\int \frac{d{\bf
k}}{(2\pi)^{3}} \left [ \frac{1}{k^2}
\dot{G}_{cc}(k_{1},t)\dot{G}_{\rho\rho}(k,t)+
\frac{1}{kk_1}\chi(k_1)\chi(k)\dot{G}_{\rho c}(k_1,t) \dot{G}_{\rho
c}(k,t)\right ] \ee
The memory kernel renormalizing the diffusion coefficient consists
of the derivatives of the correlation functions and thus vanishes in
the long time limit. The equations (\ref{nep-eqns}) obtained for the
ENE transition in the mixture are same as obtained by Harbola and
Das \cite{uh-pre} for a binary system. The NEPs show a strong
dependence on the size ratio $\alpha=\sigma_2/\sigma_1$, mass ratio
$\kappa=m_2/m_1$ and concentration of the species $x_{1}$. The value
of the total packing fraction
$\eta=\pi(N_1\sigma_1^3+N_2\sigma_2^3)/N$ at the ENE transition
point is the critical packing fraction and is denoted as $\eta_c$.
We display in Fig. \ref{fig3} how $\eta_c$, obtained from the
solution of Eq. (\ref{nep-eqns}) changes with concentration $x_1$ of
species $1$ for different values of the size ratio $\alpha$ of the
particles (with mass ratio $\kappa=1$).

\section{Comparison with the existing MCT Model} 
\label{sec5}

In this section we discuss a related model of MCT for a binary
mixture  which has been used extensively earlier in the literature
and discuss its relevance in comparison to the present work. We
briefly indicate how the method which we have used above for
obtaining our model can also obtain the existing model and clarify
the un-physical approximations made in this case. The existing
version of the MCT \cite{bosse} for binary systems is reproduced
using the present method with a different choice for the set of slow
variables. Here in addition to the individual conserved densities
\{$\rho_1$,$\rho_2$\}, the momentum densities of each component are
treated as two {\em separate} slow variables.  The microscopic
definitions for the mass and  momentum  densities are given in Eqs.
(\ref{rhos-def})-(\ref{gs-def}). The reversible parts of the
corresponding equations for the slow modes are obtained by using the
Poisson bracket relations among these four densities. However,
assuming widely separated time scales, stochastic equations are
written for each of the momentum densities ${\bf g}_s$ ($s=1,2$).
Using the same driving free energy functional $F$ as in the
Sec.~\ref{sec2} and further illustrated in Appnedix \ref{appendix1},
the generalized Langevin equation corresponding to the four ``slow"
variables $\{\rho_s,{\bf g}_s\}$ for $s=1,2$ are obtained
\cite{uh-pre}:
\bea \label{deq1}
&&\frac{\partial \rho_{s}}{\partial t}+ {\bf \nabla} \cdot {\bf g}_s =0 \\
\label{momen1} &&\frac{\partial {\rm g}_{is}}{\partial t}+\nabla_j {{\rm g}_{is}
{\rm g}_{js} \over\rho_s} +\rho_s \nabla_i \frac{\delta F_U}{\delta
\rho_s}+ L_{ij}^{ss^{\prime}} \frac{\delta F}{\delta
{\rm g}_{js^{\prime}}} ={\xi}_{is}~~. \eea
In writing Eq. (\ref{deq1}), the self and inter-diffusion of the two
species in the density equations  have been ignored. This is
equivalent to writing ${\bf g}_{s} = ({\rho_s}/{\rho}){\bf g}$. The
thermal noise ${\xi_s}$ in the equations for the momentum density
${\bf g}_s$ follows the fluctuation dissipation relation to the bare
transport coefficients,
\begin{equation}
\label{B-noise} <\xi_{is}({\bf x},t)\xi_{js^\prime} ({\bf
x}^\prime,t^\prime)> = 2k_BT {L_{ij}^{ss^\prime}} \delta({\bf
x}-{\bf x^\prime}) ~\delta(t-t^\prime)~.
\end{equation}
Renormalization of the bare transport coefficients, as a result of
the nonlinearities in the equations for the momentum currents is
computed within the self-consistent mode coupling approximations of
dominant density fluctuations. To the one loop order the
contributions to the longitudinal component of the various elements
of the transport coefficient matrix ${\cal L}_{ss^\prime}$ are
obtained as,
\be \label{omctmk} \tilde{\mathcal
L}_{ss^{\prime}}(q,t)=\frac{n_0}{2n_{s}n_{s^{\prime}}}
\int{\frac{d^{3}k}{(2\pi)^{3}}}
\sum_{\mu,\sigma,\mu^{\prime},\sigma^{\prime}}V_{s\mu\sigma} ({\bf
q},{\bf k},{\bf k_{1}})V_{s^{\prime}\mu^{\prime}\sigma^{'}} ({\bf
q},{\bf k},{\bf k_{1}})
G_{\mu\mu^{'}}(k,t)G_{\sigma\sigma^{\prime}}(k_{1},t), \ee
where ${\bf k}_{1}={\bf q}-{\bf k}$ and $n_0(=n_1+n_2)$ is the total
number density with $n_1$ and $n_2$ being the individual number
densities of species 1 and 2 respectively. The expression for the
vertex function $V_{s\mu\sigma}$ is given by
\be V_{s\mu\sigma}(q,k)=\frac{n}{m_{\mu}m_{\sigma}}[\hat{\bf q}.{\bf
k}\delta_{s\sigma}c_{s\mu}(k) +\hat{\bf q}.{\bf
k}_{1}\delta_{s\mu}c_{s\sigma}(k_{1})]. \ee
Starting from the above set of equations of motion, it is
straightforward to obtain (using the MSR theory outlined in the
Sec.~\ref{sec3}) a set of nonlinear integro-differential equations
or the so called memory function equations for the elements of the
correlation function matrix $G_{ss^\prime}$. The location of the ENE
transition in the model is obtained by considering the long time
limit of the equations for the time evolution of the correlation
functions. The dynamical transition point in the previous version of
MCT model is located by using a matrix equation similar to Eq.
(\ref{nep-eqns}).
\be \label{nepb}F(q)=\frac{1}{q^2}{\mathcal S}(q){\mathcal
L}(q)[{\mathcal S}(q)-F(q)], \ee
where the matrix related to the structure is defined as ${\mathcal
S}(q)=\sqrt{x_{i}x_{j}}S_{ij}$ and ${\mathcal L}_{ss^{\prime}}(q)$
is the long-time limit of the memory function ${\mathcal
L}_{ss^{\prime}}(q,t)$. These integral equations for the NEPs are
similar to the equations we obtain in our model discussed in the
earlier section. The actual form of the integral equations in the
respective cases are determined from the wave-vector dependent
structure factors or the equilibrium correlation functions which are
used as an input in the model. These structural inputs are
determined by the driving free energy functional $F$ for the system.
In this respect it is useful to note that {\it same free energy
functional or wave-vector dependence} of the equilibrium
correlations are used here for all the models. In the present model
the ENE transitions corresponding to various choices for the
thermodynamic parameters for mixture occur at higher packing
fraction values than that predicted from Eqs. (\ref{nepb}). This
also agrees well with the results seen in the computer simulations
on binary systems \cite{nauroth,nauroth-kob}. Details of such
differences between the two types of mode coupling models have
already been reported in Ref. \cite{uh-pre}.

\begin{figure}[h]
\includegraphics[height=10cm]{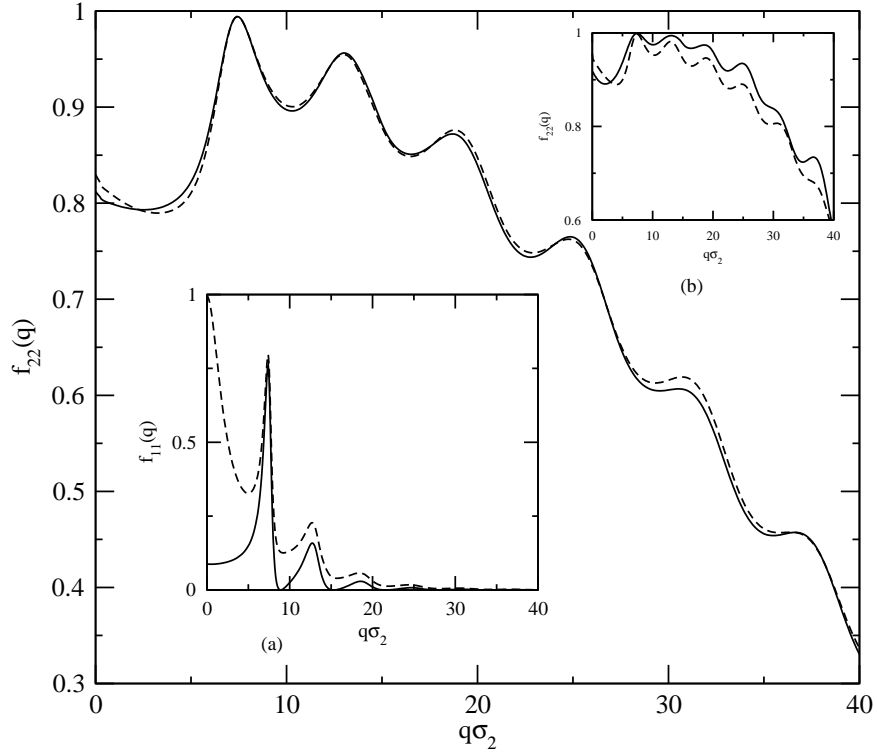}
\caption{ Comparison between existing MCT model(mass ratio
independent) and the present model in the Brownian limit of large
mass ratio $\kappa=m_2/m_1=10^4$. For a mixture with size ratio
$\alpha=\sigma_2/\sigma_1=10^2$, packing fraction $\eta=0.6$, and
for $x_2=.1$, the Non ergodicity parameters $f_{22}(q)$ (Main
figure) and $f_{11}(q)$ (Inset a)  vs. $q\sigma_2$. Inset (b) is
same as Inset (a) for $x_2=.01$. For all three figures, solid lines
are results from Eq. (\ref{nep-eqns}) of present work and dashed
lines are from Eqs. (\ref{nepb}) of Ref. \cite{bosse} of existing
MCT.} \label{fig4}
\end{figure}

\begin{figure}[h]
\includegraphics[height=10cm]{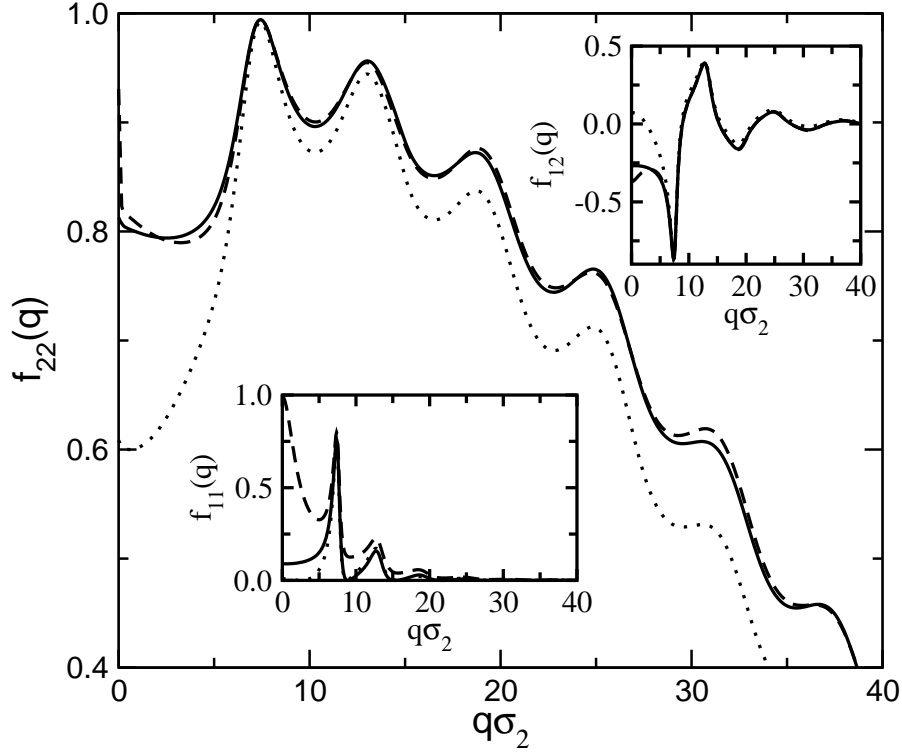}
\caption{Comparison between existing MCT model and the present model
in the Brownian limit of large mass ratio. Results are for the same
mixture as in Fig. \ref{fig4} with size ratio
$\alpha=\sigma_2/\sigma_1=10^2$, packing fraction $\eta=0.6$, and
for $x_2=.1$. The Non ergodicity parameters $f_{22}(q)$ (Main
figure) and $f_{11}(q)$ (lower Inset) and $f_{12}(q)$ (upper inset)
vs. $q\sigma_2$. For all three figures, results from Eq.
(\ref{nep-eqns}) of present work for mass ratio
$\kappa=m_2/m_1=10^6$ (solid line); $=10^2$ (dotted line), and from
Eqs. (\ref{nepb}) of Ref. \cite{bosse} of existing MCT(dashed
line).} \label{fig5}
\end{figure}

In the present section we focus on the limiting case in which the
previous MCT models \cite{bosse} agree with predictions of our work
with respect to the location of the ENE transition. It is clear from
the deductions presented above that the primary difference between
the two models come from the treatment of the momentum densities for
the two different species of the mixture. The individual momentum
densities $\{{\bf g}_1,{\bf g}_2\}$ are not conserved variables and
there is no physical basis in assuming a separation of slow and fast
time scales in their dynamics. However, existing mode coupling model
is obtained using separate Langevin equations for each of the
momentum densities $\{{\bf g}_1,{\bf g}_2\}$. On the other hand the
total density ${\bf g}={\bf g}_1+{\bf g}_2$ is a conserved mode and
has been treated as a slow mode in our model. For the Brownian
particle, it is appropriate to write a Langevin equation for the
momentum density of the single particle with high inertia. For the
coarse grained tagged particle, the momentum density ${\bf g}^B({\bf
x},t)$ and the corresponding mass density $\rho^B({\bf x},t)$ are
related by the continuity equation,
\be \label{rb-eq} \frac{\partial\rho^B}{\partial t}+ \nabla . {\bf
g^B}=0~~. \ee
For tagged particle momentum density ${\bf g}^B({\bf x},t)$, the
following equation is written down
\begin{eqnarray}
\label{gb-eq} \frac{\partial {\rm g}^B_{i}}{\partial t}+
\rho^B\nabla_i\frac{\delta F}{\delta \rho^B}+ \int d{\bf x}^\prime
\Gamma^B_0 \frac{\delta F}{\delta {\rm g}^B_i}=f^B_i~.
\end{eqnarray}
The bare friction coefficient $\Gamma^B_0$ is related to the noise
${\bf f}^B$ through the usual fluctuation dissipation relation. For
the collective density $\rho$ and current ${\bf g}$, the
corresponding equations of motion follow from microscopic
conservation laws; it is not so for ${\bf g}^B$. In this case the
total momentum density ${\bf g}$ is approximated well by the rest of
the ``mixture". Therefore {\em only} for the case of a Brownian
particle the set of four equations for the conserved densities
(\ref{deq1}-(\ref{momen1}) can be identified respectively with the
set (\ref{cont-eqn})-(\ref{momt-eqn}) and
(\ref{rb-eq})-(\ref{gb-eq}) in our present model. Hence in this case
the results obtained in the existing MCT model comes close to the
present formulation which keeps a proper account of the conservation
laws. The mass ratio dependence of the ENE transition with the NEP
Eqs. (\ref{nep-eqns}) and the wave vector dependence of the
corresponding NEPs has been reported in detail in an earlier work
\cite{uh-jsp}. In the present section we demonstrate this
equivalence of the two MCT models for the case of a single Brownian
particle (large inertia) in a simple liquid being treated as a
mixture. Using a very large mass ratio and very small concentration
of the heavier particle, {\em i.e.}, $\kappa{\rightarrow}$large and
$x{\rightarrow}$small ( signifying the Brownian limit ) the results
for the NEP in the present model for most wave vector values are
very close to the predictions of the existing model. This is shown
in Fig. \ref{fig4}. In the inset where $f_{11}$ is shown, the $q\to
0$ behavior of the two models are very different. To see this we
need to take into account the behavior of the  tagged-particle
correlation $\phi_s$ in the two models. Taking the form of the
Laplace transform $\phi_s(q,z)$ in the form of a diffusive pole with
generalized diffusion coefficient $D_s(q,z)$, the long time limit of
the correlation function or the so called nonergodicity parameter
$f_s$ is obtained as
\be \label{neps} \frac{f_{s}(q)}{1-f_s(q)}= \frac{z}{iq^2D_s(q,z)}.
\ee
Since for the existing MCT models, self-diffusion  constant vanishes
in the small frequency limit as $D_s\sim{z}$, we note from Eq.
(\ref{neps}) that the non ergodicity parameter $f_s(q)$ gets pinned
at the value $1$ in the same limit. However, in our model $D_s$ is
finite in $z\to 0$ limit, {\em i.e.}, $z/D_s$ for a fixed $q$ is
vanishing in this limit. The relation (\ref{neps}) implies vanishing
of the nonergodicity parameter $f_s$ for small wave numbers. For the
exisiting MCT model, the predictions for the dynamics are in fact,
independent of mass ratio. In Fig. \ref{fig5} we show another
comparison of the two models at the same packing fraction $\eta=0.6$
for a mixture having size ratio $\alpha=10^2$  and $x_2=0.1$. The
two mass ratios used here are $\kappa = 10^2$ and $10^4$
respectively. Therefore though in the Brownian limit, the two MCT
models are approximately matching, differences with our model show
up at small $q$ due to role of the conservation laws. For the other
component $f_{22}$ both models essentially represent the collective
correlations for a homogeneous liquid and hence they agree. However
positivity of self-diffusion for interacting Brownian particles with
hard core has been rigorously demonstrated \cite{osada}.

\section{Single particle dynamics}
\label{sec6}

We now consider the implications of the MCT developed here on the
dynamics of a tagged particle in a sea of identical particles. In
this non-perturbative analysis we follow a method developed recently
in Ref. \cite{DM09} to establish, with the use of the available
fluctuation dissipation relations in the MSR field theory, the long
time behavior of time correlation functions. This was developed for
analyzing the asymptotic dynamics of the correlation functions for
the collective variables in a one component fluid. Here we apply
this method for the binary mixture to prove an ENE transition beyond
a critical density. Furthermore, the $1/\rho$ nonlinearities in the
dynamics are ignored in this case. We then consider the model in the
so called one component limit (to be explained below) to study the
nature of a tagged particle motion in a homogeneous liquid. This
analysis demonstrates the decoupling between the collective and
single particle dynamics in a dense liquid.

\subsection{The ENE transition in the mixture}

We consider Eq. (\ref{eq:84}) for the correlation function for the
MSR theory outlined above for a binary system. The ENE transition is
characterized by the density and concentration  correlation
functions
$\{G_{\rho\rho}(\omega),G_{\rho{c}}(\omega),G_{cc}(\omega)\}$ each
developing a $\delta$-function contribution. Using the one loop
results (\ref{oneloop-L})-(\ref{oneloop-D}) respectively for the
corresponding memory functions, we make the following observations:

\noindent A. The generalized transport coefficient $L(\omega)$,
which is the Laplace transform of $L({\bf q},t)$ defined in Eq.
(\ref{oneloop-L}) involves  the correlation  of the $\rho$ and $c$.
Hence at the ENE transition, $L(\omega)$ has a singular part with
$\delta(\omega)$ contribution. This conforms to the physics of the
viscosity of the mixture diverging in the ideal glass phase.
Equivalently that the self-energy $\Sigma_{\hat{g}\hat{g}}$ blows up
at small frequencies and hence is written with a general
non-perturbative expression:
\begin{equation} \Sigma_{\hat{g}\hat{g}} =-A\delta (\omega ) +
\Sigma^{R}_{\hat{g}\hat{g}}~. \label{eq:22}
\end{equation}
The second term in the RHS represents parts of the self-energy
contribution which are regular in the $\omega\rightarrow{0}$ limit.
In writing the above expression wave-vector dependence in the model
is not ignored but suppressed to keep the notation simple.

\noindent B. From Eq. (\ref{oneloop-D}) since it follows that the
renormalization of the $\nu(q,z)$ involves only derivatives of the
correlations of $\rho$ and $c$, it has no singular contribution
($\sim \delta(\omega)$) of $\nu(\omega)$ or $\nu^{\prime}(\omega)$
in the small $\omega$ limit.

\noindent To test compatibility with the Dyson equation
corresponding to the MSR action (\ref{Aaction-MSR}), we substitute
Eq. (\ref{eq:22}) back into Eq. (\ref{eq:84}). This involves
setting both $\alpha$ and $\beta$ equal to $\rho$ in Eq.
(\ref{eq:84}). It is straightforward to obtain that the singular
contribution of $G_{\rho\rho}$ comes from that in the self energy
$\Sigma_{\hat{g}\hat{g}}$ in the form,
\begin{equation}
\label{eq:23} G_{\rho\rho} = -A G_{\rho\hat{g}}\delta(\omega)
G_{\hat{g}\rho} + \bar{\Sigma}^R_{\hat{g}\hat{g}}~,
\end{equation}
where $\bar{\Sigma}^R$ is the part of the correlation function
contributed by the regular part $\Sigma^{R}$. For an ENE transition
to occur it is needed that the response function $G_{\rho\hat{g}}$
does not vanish as $\omega\rightarrow 0$. The response functions
$G_{\alpha\hat{\beta}}$ are calculated from Eq. (\ref{resp-matx}),
where $N_{\alpha\hat{\beta}}$ are as given in Table \ref{table3}.
\begin{equation}
\label{resp1} G_{\rho\hat{g}} = \frac{N_{\rho\hat{g}}}{\cal
D}=\frac{q(\omega+iq^2\nu ~)}{\cal D} ~~.
\end{equation}
The right hand side of the above equation is nonzero since
$\nu(\omega)$ and $\nu^\prime(\omega)$, as defined in Eqs.
(\ref{renp-mu}) and (\ref{renp-mup}) respectively both are nonzero
in the $\omega\rightarrow 0$ limit. In the same zero frequency
limit, the determinant ${\cal D}$ do not blow up as $\omega{L}$,
$c^{2}$ and $\nu$ are finite. With these assumptions, ${\cal
D}(\omega\rightarrow{0})$ is not infinite and hence
$G_{\rho\hat{g}}{\ne}0$ in the low frequency limit. Therefore the
density correlation function $G_{\rho\rho}$ develops a
$\delta(\omega)$ part. In an exactly similar way it follows that the
correlations $G_{\rho{c}}$ and $G_{cc}$ each develop a singular part
($\sim\delta(\omega)$ ) by coupling to $\Sigma_{\hat{g}\hat{g}}$.
This is a consequence of the fact that both $\nu(\omega)$ as well as
$\nu^\prime(\omega)$ are nonzero in the small $\omega$ limit.

In comparison to the above result, the correlation functions
involving a momentum index $\mathrm{\bf g}$ do not contain a
$\delta$-function peak at zero frequency. To demonstrate this, we
note that if either of the indices $\alpha$ or $\beta$ in the left
hand side of Eq. (\ref{eq:84}) is set equal to $\mathrm{g}$ then the
singular contribution in $\Sigma_{\hat{g}\hat{g}}$ is coupled to the
response function $G_{g\hat{g}}$. From Table \ref{table3}, it
follows that
\begin{equation}
G_{g\hat{g}}=\frac{\omega (\omega+iq^2\nu ~)}{{\cal D}(\omega)}~,
\label{gg-resp}
\end{equation}
which means that $G_{g\hat{g}}$  vanishes as $\omega\rightarrow 0$
as long as ${\cal D}(\omega =0)\neq 0$. Therefore the correlation
functions involving a momentum index $\mathrm{g}$ do not show a
$\delta$-function peak at zero frequency.

To summarize, for a binary mixture all the three correlations
$G_{\rho\rho}(\omega)$, $G_{\rho{c}}(\omega)$, and $G_{cc}(\omega)$
develop a singular piece $\sim{\delta}(\omega)$ or equivalently
develop a nonzero long time limit signifying an ENE transition. It
is important to note here that the possibility of the ENE transition
requires that both $\nu(\omega)$ as well as $\nu^\prime(\omega)$ are
nonzero in the small $\omega$ limit.

\subsection{Tagged particle dynamics}

We consider the system for which the two species are identical, {\em
i.e.}, the size ratio $\alpha$ and mass ratio $\kappa$ are both
equal to 1 and the number of the particles $N_1=1$ and $N_2=N-1$.
This will be referred to as the one component limit in the
following. For large $N$, the relative fractions for the two species
are $x_{1}\rightarrow 0$ and $x_{2}\rightarrow 1$. The concentration
variable $c$, defined in Eq. (\ref{conc-def}) reduces to the tagged
particle density $\rho_{s}$ and hence the correlation function
$G_{cc}$ reduces to tagged particle correlation function $G_s$
\cite{boon-yip}. The present theory therefore reduces to the MCT for
the total and tagged particle dynamics in a one component liquid. We
first consider the behavior of the corresponding static
susceptibilities $\chi_{\rho\rho}$, $\chi_{cc}$ and $\chi_{\rho c}$.
The inverse static-susceptibilities are respectively obtained in the
one component limit as
\be \label{stf-ocl} \beta\chi_{\rho\rho}^{-1}({\bf k})=
\frac{1}{m^2nS(k)}~,~~~ \beta\chi_{cc}^{-1}({\bf k}) =
\frac{1}{m^2n{x_1}{x_2}}~,~~~ \beta\chi_{\rho c}^{-1}({\bf k}) = 0~.
\ee
The above results also agree with the wave-vector dependent formulas
\cite{RY} obtained by using Ornstein-Zernike relations
\cite{frisch,lebowitz,ashcroft-leng} for the partial structure
factors of a binary liquid. The key point here is that the off
diagonal-element ($\rho-{c}$) vanishes and the susceptibility matrix
is diagonal in the one component limit. From the definition
(\ref{renp-mup}) it directly follows that $\nu^\prime(\omega)$ which
is nonzero for a binary mixture, vanishes in this limit. As a result
of this, we obtain $N_{c\hat{g}}\rightarrow 0$. The above analysis
implies that the correlation function $G_{cc}$ does not have any
contribution coming from the singular part ($\sim \delta(\omega)$)
in the self energy $\Sigma_{\hat{g}\hat{g}}$. Hence $G_s\rightarrow
0$ in the small frequency limit. However with $\nu^{\prime}$ ( and
hence $\nu_1$) vanishing, we have
\begin{equation}
G_{\rho\hat{g}} = \frac{q}{\omega^2 -q^{2}c^{2}+iq^2\omega L}~.
\label{resp1-ocl}
\end{equation}
Thus $G_{\rho\hat{g}}$ is nonzero in the small $\omega$ limit.
Applying the same argument used above with Eq. (\ref{eq:23}), the
collective density correlation function $G_{\rho\rho}$ has the
singular contribution from $\Sigma_{\hat{g}\hat{g}}$ self-energy.
Thus in the one component limit the collective density correlation
function freezes at the ENE transition although the tagged particle
correlation $G_s(q,t)$ goes to zero in long time limit. The
corresponding self-diffusion coefficient for the tagged particle is
nonzero in the hydrodynamic limit. Thus, the single particle
dynamics decouples from the dynamics of collective variables. In
this respect the conclusion of the present work differs
fundamentally from existing mode coupling model for binary systems.

\section{Discussion}
\label{sec7}

We have studied here the mode coupling dynamics of a binary mixture
in terms of the microscopically conserved densities for the two
component system, namely the total density $\rho$, the concentration
variable $c$, and the total momentum density ${\bf g}$. The dynamics
is described in terms of nonlinear Langevin equations for the modes
with white noise. The correlations of the noise in the respective
stochastic equations of motions, define the bare transport
coefficients for the system. In the present formulation keeping
consistency with the white noise, there are two transport
coefficients, namely the bare viscosity $L^0$ and the $\gamma_0$ for
the inter-diffusion. Using a MSR field theory we have studied the
effects of the nonlinearities in the Langevin equations of the slow
variables on the long time limit of dynamic correlations. From a
careful consideration of the renormalizability of the theory, we
obtain the relevant mode coupling contributions which drive the
system to an ENE transition. It is shown respectively in Eqs.
(\ref{renorm-nu}) and (\ref{renorm-nup}) that the renormalized
inter-diffusion $\gamma_0$ couples through the respective static
correlations $\chi^{-1}_{c c}$ and $\chi^{-1}_{\rho{c}}$. Two
effective transport coefficients $\nu$ and $\nu^\prime$ appear and
for the mixture of two species, with finite concentrations of each,
both are nonzero. For such a mixture, we have shown that the
feedback mechanism from slowly decaying density fluctuations drives
the system in to an ENE transition at which the all the three
correlation functions $\{G_{\rho\rho},G_{\rho{c}},G_{cc}\}$ freeze
in the long time limit.

We also demonstrate here the conditions in which our results agree
with existing formulation of MCT for a mixture. In the existing
theory the individual momentum densities are treated as slow
variables with separation of time scales in its dynamics, though
these are non conserved modes. For the Brownian limit (with high
inertia) however it is appropriate to write a Langevin equation for
the momentum density of the single particle. A key aspect of our
formulation of the MCT is that the location of the ENE transition as
well as the nature of the dynamic correlation is now dependent on
the mass ratio of the constituent particles of the mixture
\cite{uh-jsp}. This is also in agreement with computer simulation
results \cite{prl03}. Study of our model with large mass ratio and
very low concentration of one of the species shows that its results
agree closely with the corresponding prediction of the NEP in the
existing MCT model.
The distinct nature of the single particle dynamics in a sea of {\em
identical} particles, observed here, is primarily a consequence of
taking into account the  conservation laws properly in the present
theory. The physics involved is very different from that for the
situation where a tagged particle diffusion  differs from collective
dynamics in a very asymmetric mixtures with sufficient
size-disparities \cite{voigtmann} of the constituent particles.
Hence it is more a geometrical effect and is linked to the
peculiarities of the physics of cage formation process in such
mixtures having very dissimilar components. Finally, though somewhat
speculative at this point, it is useful to note here that the
observed decoupling of collective dynamics from single particle
motion, is more indicative of the violation of the Stokes Einstein
relation \cite{biroli-jpcm} than the converse. In existing MCT,
since divergences are driven by that of the relaxation time of the
density correlations, Stokes-Einstein relation is not violated.

Dynamical light-scattering experiments for a bi-disperse mixture of
colloidal particles \cite{williams} with size-ratio 0.6 interacting
nearly by a hard-sphere potential indicate that, increasing the
concentration of smaller particles beyond a certain value slows down
the dynamics in the system. This so called plasticization effect,
which has been observed in existing MCT \cite{voigtmann}, is also
confirmed from our model equations. In Fig. \ref{fig3}, the plot of
the critical packing fraction $\eta_c$ , {\em i.e.}, the minimum
packing fraction at which the binary mixture undergoes an ENE
transition for a particular size-ratio and concentration , against
the concentration of smaller particles $x_1$ is displayed for  five
different size-ratios. The plasticization effect to hold for all
values of the size-ratio. The effect is however more pronounced for
mixtures with high size-disparity.

The present formulation of MCT reduces to the dynamics of both
collective as well as the tagged particle density correlation
functions for a one component system. In the final section of the
paper we have considered this so called one component limit of our
model. Our analysis demonstrates that the dynamics of the tagged
particle correlation and the total density correlation are decoupled
in this limit. The role of the static correlations is vital here. In
the one component limit $\nu^\prime$ becomes zero due to the
vanishing of the susceptibility factor in Eq. (\ref{renorm-nup}). In
this case we show that the collective correlation freezes at the
transition while the single particle correlation decays to zero.
Hence the self diffusion coefficient remains finite. In the existing
MCT model for one component systems \cite{beng}, with the same
static correlation matrix, the tagged particle dynamics is slaved to
that of the total density correlation. According to these theories,
at the ENE transition of the MCT, both time auto-correlation
functions, collective as well as single particle are simultaneously
nonzero in the long time limit. This implies that the self-diffusion
coefficient vanishes at the ENE transition. This is where our result
is crucially different from existing MCT for one component system.
In this regard it would be useful to reexamine the practice of
``locating" the ENE transition point in simulation studies or
experiments are by {\em extrapolating} the diffusion constant of a
tagged particle to zero. This so called MCT transition point does
not agree with the same obtained from the integral equations of MCT
\cite{nauroth} using equilibrium structure factor as an input.

\newpage

\appendix

\section{Structure and Dynamics of the mixture}
\label{appendix1}

In this appendix we provide a brief description of the structure and
dynamics of the binary mixture in terms of the hydrodynamic modes in
the system. The first involves the free energy functional in terms
of which equilibrium correlation are defined. The dynamics is
described with stochastic nonlinear equations of generalized
hydrodynamics.

\subsection{Free energy functional}

The equal time correlations of the hydrodynamic modes and the
structural properties of the mixture are obtained from the averages
taken in terms of the equilibrium probability distribution for the
fluctuations. See Eq. (\ref{eqt-av}) for definition. For this the
corresponding free energy functional expressed in terms of the slow
modes is necessary. In the construction of the generalized Langevin
equation (\ref{glgvn}), the the free energy $F$ as a functional of
the slow modes $\{\psi_\alpha\}$ is also needed. The functional $F$
is generally divided into two parts: the momentum density dependent
part is generally referred to as the kinetic part $F_K$ and the rest
as the potential part $F_U$:
\be \label{ftotal} F=F_K+F_U~~. \ee
The kinetic part $F_K$ is given in Eq. (\ref{fekin}). The potential
part of the free energy functional $F_U$ dependent on $\rho$ and
$c$, is taken to be Gaussian in these variables at the simplest
level in the present theory. The so called potential part
$F_U[\rho,c]$ has two contributions,
\be \label{FUP} F_U[\rho,c]=F_\mathrm{id}+F_\mathrm{in}
\ee
with $F_\mathrm{id}$ is the so called ideal gas part for a non
interacting system. In terms of the slow variables
$\{\rho_{1},\rho_{2}\}$, the ideal gas part is obtained as,
\be \label{fide} F_\mathrm{id}=\sum_{s,s^{\prime}}\frac{1}{m_{s}}
\int{dx\rho_{s}({\bf x}) \left[ {\rm ln}\left(\frac{\rho_{s}}
{\rho_{0s}}\right)-1 \right]} ~~. \ee
The ideal part of free energy functional is non-Gaussian. However,
the ${\rm log}$ term is approximated to have a Gaussian form.
$F_\mathrm{in}$ is the interaction part. The standard density
functional expansion \cite{RY} of the interaction part of the free
energy $F_\mathrm{in}$ is obtained in terms of a functional Taylor
series expansion involving the direct correlation functions
$c_{ss'}$ where $s,s'=1,2$ for the mixture.
\begin{eqnarray}
\label{Fex-RY} F_\mathrm{ex}= F_0-
\sum_{s,s'=1}^2\frac{1}{2m_sm_{s'}}\int d {\bf x}_1 \int d {\bf x}_2
c_{ss'} ({\bf x}_1,{\bf x}_2 ;n_0) \delta \rho_s({\bf x}_1) \delta
\rho_{s'}({\bf x}_2) + ..~~~,
\end{eqnarray}
where $F_0$ is the free energy of the uniform liquid state. The
direct correlation functions $c_{ss'}$ is defined in the density
functional formulation as,
\be \label{dirc-def} c_{ss'}({\bf x},{\bf x}')=
\frac{\delta^2{F_\mathrm{in}[n({\bf x})]}}{{\delta{n}_s({\bf
x})}{\delta{n}_{s'}({\bf x}')}} {\Big |}_0 ~~~, \ee
with the ``0" implying that the functional derivative is evaluated
for the equilibrium liquid state. In terms of the fields
$\{\rho,c\}$ the functional $F_U$ is written in the Gaussian form
\be \label{fepot} F_{U}[\rho,c] = \frac{1}{2}\sum_{\alpha,\sigma}
\int{d{\bf x}}\int{d{\bf x}^{\prime}} \delta\rho_\alpha({\bf
x})\chi_{\alpha\sigma}^{-1} ({\bf x}-{\bf
x}^{\prime})\delta\rho_{\sigma}({\bf x}^{\prime})  ~~. \ee
In the above expression for the free energy the
$\chi^{-1}_{\alpha\sigma}$ denotes the $\alpha\sigma$-th element of
the inverse of the equal time correlation matrix or the so called
susceptibility matrix ${\chi}_{\alpha\sigma}$ where
$\alpha,\sigma\in\{\rho,c\}$. The above free energy is also
conveniently expressed as a Gaussian functional of the pair
$\{\rho_1,\rho_2 \}$ instead of the set $\{\rho,c\}$ as,
\be F_{U}[\rho_s] = \frac{1}{2}\int{d{\bf x}}\int{d{\bf x}^{\prime}}
\sum_{s,s'=1}^2 \delta\rho_s({\bf x})\chi_{ss'}^{-1} ({\bf x}-{\bf
x}^{\prime})\delta\rho_{s'}({\bf x}^{\prime}) \label{fepot12} ~~.
\ee
Using the results (\ref{FUP}), (\ref{fide}), and (\ref{dirc-def}),
and doing a quadratic order expansion in density fluctuations, the
$\chi_{ss'}$'s are expressed in terms the corresponding direct
correlation functions $c_{ss'}$ for a mixture through
Ornstein-Zernike relations \cite{frisch,ashcroft-leng}. The elements
of the static susceptibility matrix $\chi_{\alpha\sigma}(q)$ are
obtained in terms of the direct correlation functions
$c_{{\alpha}\sigma}$ with $\alpha,\sigma\in\{\rho,c\}$ as follows:
\begin{eqnarray}
\label{d-chi-rr}
\chi_{\rho\rho}^{-1}(q)&=&\frac{\beta^{-1}}{m_1^2n_0}~
\left[x_1+\frac{x_2}{\kappa^2}-c_{\rho\rho}(q)\right],\\
\label{d-chi-rc}\chi_{\rho
c}^{-1}(q)&=&\frac{\beta^{-1}}{m_1m_2n_0}~
\left[\kappa-\frac{1}{\kappa}-c_{\rho c}(q)\right],\\
\label{d-chi-cc}\chi_{cc}^{-1}(q)&=&\frac{\beta^{-1}}{m_2^2n_0}~
\left[\frac{\kappa^2}{x_1}+\frac{1}{x_2}-c_{cc}(q)\right]~~,
\end{eqnarray}
where $c_{\rho\rho}(q)$, $c_{\rho c}(q)$ and $c_{cc}(q)$ are given
by the following expressions :
\begin{eqnarray}
\label{d-c-rr}c_{\rho\rho}(q)&=&x_{1}^{2}
\bar{c}_{11}(q)+2(x_{1}x_{2}/\kappa)\bar{c}_{12}(q)
+(x_{2}^{2}/\kappa^2)\bar{c}_{22}(q),\\
\label{d-c-rc}c_{\rho
c}(q)&=&x_{1}\kappa\bar{c}_{11}(q)+(x_{2}-x_{1})
\bar{c}_{12}(q)-(x_{2}/\kappa)\bar{c}_{22}(q),\\
\label{d-c-cc}
c_{cc}(q)&=&\kappa^2\bar{c}_{11}(q)-2\kappa\bar{c}_{12}(q)+\bar{c}_{22}(q).\
\end{eqnarray}
We have used in Eq. (\ref{d-c-rr}) the notation $\bar
c_{ss'}(q)=n_0c_{ss'}(q)$ for $s=1,2$. The quantity $\kappa$ is the
mass ratio $m_2/m_1$ of the two species and the relative abundance
is denoted as $x_s=N_s/N$ for $s=1,2$.

\subsection{The Generalized Langevin equations}

The Langevin equation (\ref{glgvn}) involves a deterministic or slow
part expressed in terms of the variables $\psi_\alpha$ and a
stochastic part. These are respectively given by a) the first two
terms on the RHS and b) the third term on the RHS. The calculation
of reversible part of the Langevin equations for the slow variables
requires the Poisson bracket relations between the slow variables.

\subsubsection{Poisson Brackets}

Using the basic Poisson bracket relations between the phase space
coordinates and the microscopic Eqs.
(\ref{rhos-def})-(\ref{conc-def})
 for the slow variables $\rho({\bf x,t})$, $c({\bf
x},t)$ and ${\rm g}({\bf x},t)$ we obtain,
\bea \lbl{prg} \{\rho({\rm x}), {\rm g}_{i}({\bf x}^{\prime})\} &=&
-\nabla_{i}[\delta({\bf x}-{\bf x}^{\prime})\rho({\bf x})],\\
\lbl{pcg} \{c({\bf x}), {\rm g}_{i}({\bf x}^{\prime})\}
&=&-\nabla_{i}[\delta({\bf x}-{\bf x}^{\prime})c({\bf x})],\\
\lbl{pgg}\{{\rm g}_{i}({\bf x}), {\rm g}_{j}({\bf x}^{\prime})\} &=&
-\nabla_{j}[\delta({\bf x}-{\bf x}^{\prime}) {\rm g}_{i}({\bf
x})]+\nabla_{i}^{\prime} [\delta({\bf x}-{\bf x}^{\prime}){\rm
g}_{j}({\bf x}^{\prime})]. \eea
All other Poisson brackets between the different members of the set
$\{\rho,c,{\bf g}\}$ are zero.

\subsubsection{Bare Dissipation Coefficients}

Next we consider the dissipative and stochastic contributions
respectively  given by the second and third terms in the RHS of Eq.
(\ref{glgvn}). The various elements of the matrix
$\Gamma_{\alpha\sigma}$ are chosen keeping consistency with the
structure of the Langevin equations. The continuity equation is
maintained for the density equation. For the density field
$\rho({\bf x},t)$ and the corresponding current, {\it i.e.}, the
momentum density field ${\bf g}({\bf x},t)$ is conserved,
 all the dissipative terms involving $\rho$ {\it viz.}
$\Gamma_{\rho\psi}=\Gamma_{\psi\rho}=0.$ For determining the
elements of the noise correlation matrix between ${\rm g}$ and $c$,
the symmetry considerations and compatibility with the white noise
both play important roles. In this respect we note that the
dissipative tensor $\Gamma_{\alpha\sigma}$ should follow the time
reversal symmetry given by
\begin{equation}
\Gamma_{\alpha\sigma}(-t)=\epsilon_{\alpha}\epsilon_{\sigma}
\Gamma_{\alpha\sigma}(t),
\end{equation}
where $\epsilon_\alpha=\pm 1$ represents the time reversal signature
of the slow variable $\psi_\alpha$. Applying this to the diagonal
elements $\Gamma_{cc}$ and $\Gamma_{{\rm g}_{i}{\rm g}_{j}}$ we
obtain $\Gamma_{cc}(-t)=\Gamma_{cc}(t)$ and $\Gamma_{{\rm g}_{i}{\rm
g}_{j}}(-t) = \Gamma_{{\rm g}_{i}{\rm g}_{j}}(t)$. However, for the
element $\Gamma_{c{\rm g}_i}$ the implications are different:
\be \label{dis-cg} \Gamma_{c{\rm g}_{i}}(-t) = -\Gamma_{c{\rm
g}_{i}}(t). \ee
The element $\Gamma_{c{\rm g}_{i}}$ is related to the correlations
of noise $f$ (say) in the $c$-equation  and that in the ${\bf
g}$-equation, {\em i.e.}, ${\bf \theta}$ through a
fluctuation-dissipation relation
\begin{equation}
\left\langle f({\bf x},t)\theta_{i}({\bf x}^{\prime},t^{\prime})
\right\rangle=2{\beta^{-1}} \Gamma_{c{\rm g}_{i}}
\delta(t-t^{\prime})\delta({\bf x}-{\bf x}^{\prime}).
\end{equation}
We note here that the symmetry (\ref{dis-cg}), cannot be maintained
if $\Gamma_{\alpha\sigma}(t)$ represents white noise, {\em i.e.}, is
represented by a delta function. However, for construction of the
FNH equations and validity of the formalism adopted here the white
noise is an essential input. For the conserved modes considered
here, the separation of time scales is valid. Therefore to maintain
consistency with the white noise we  take $\Gamma_{c{\rm
g}_{i}}=\Gamma_{{\rm g}_{i}c}=0$. The noise correlation in the $c$
and ${\bf g}$ equations are given by fluctuation dissipation
relations:
\bea \lbl{cnoise-cor2} \left\langle f({\bf x},t)f({\bf
x}^{\prime},t^{\prime}) \right\rangle &=& 2{\beta^{-1}}\gamma_{cc}
\delta({\bf x}-{\bf x}^{\prime})\delta(t-t^{\prime}) \\
\lbl{gnoise-co1r} \left\langle \theta_{i}({\bf x},t)\theta_{j}({\bf
x}^{\prime},t^{\prime}) \right\rangle &=& 2{\beta^{-1}}L^0_{ij}
\delta({\bf x}-{\bf x}^{\prime})\delta(t-t^{\prime}),\\
\lbl{cgnoise-cor1} \left\langle f({\bf x},t) \theta_i ({\bf
x}^{\prime},t^{\prime}) \right\rangle &=& 0. \eea
For the diagonal element $\Gamma_{cc}$, for keeping consistency with
conservation laws, we have $\Gamma_{cc} = \gamma_{0}\nabla^2$. On
the other hand $\Gamma_{{\rm g}_i{\rm g}_j}$ elements which appear
in the momentum equation are represented in terms of the bare
viscosities $\Gamma_{{\rm g}_i{\rm g}_j}=L^0_{ij}$. For an isotropic
system the $L^0_{ij}$ matrix is described in terms of two
independent quantities, namely the bulk viscosity $\zeta_0$ and the
shear viscosity $\eta_0$ are defined as:
\be \label{visc-tensor1} L^0_{ij}= (\zeta_0 + \frac{\eta_0}{3})
\nabla_{i}\nabla_{j} +\eta_0 \delta_{ij}{\nabla}^2 ~~. \ee
The longitudinal viscosity is defined as $L_0=\zeta_0+4\eta_0/3$.
Taking into account the Poisson bracket relations given by Eqs.
(\ref{prg})-(\ref{pgg}), and the Eqs. (\ref{fekin}), (\ref{fepot}),
and (\ref{ftotal}) for the free energy functional $F$ we obtain the
following Langevin equations respectively for the set of slow
variables $\{\rho,c, {\bf g}\}$ stated in Eqs.
(\ref{cont-eqn})-(\ref{conc-eqn}).

\section{The Martin-Siggia-Rose field theory}

\label{appendix2}

\subsection{The MSR action functional}

In the standard Martin-Siggia-Rose (MSR) formalism, the correlation
and response functions are determined using an action functional
constructed for the field theory. For a set of fields
$\{\psi_\alpha\}$ with equations of motion given by (\ref{glgvn}),
the average of a functional $f[\psi]$ is obtained as \cite{mybook}
\begin{eqnarray}
\label{aver-31}  <f[\psi]> = \frac{ \int D\psi \int D\hat{\psi}
f[\psi] \exp \left [ -{\cal A}[\psi,\hat{\psi}] \right ]} {  \int
D\psi \int D\hat{\psi} \exp \left [ -{\cal A}[\psi,\hat{\psi}]
\right ]}~~.
\end{eqnarray}
The action functional ${\cal A}[\psi,\hat{\psi}]$ is obtained as
\begin{eqnarray}
\label{msr-ac4} {\cal A}[\psi,\hat{\psi}] = \int d1 \int d2
\hat{\psi}(1)\beta^{-1}\Gamma^0(12)\hat{\psi}(2) + i \int d1
\hat{\psi}(1)\left \{ {{\partial \psi(1)} \over {\partial t_1}} +
\left [ Q+\Gamma^0 \right ] \frac{\delta F}{\delta\psi} \right \}~~,
\nonumber
\end{eqnarray}
where we have not explicitly written the the field indices to avoid
cluttering. The expression (\ref{aver-31}) for the average of the
functional $f({\psi})$ is used to write the averages of fields and
higher order correlation functions in terms of a generating
functional $Z_\xi$. Assuming $f(\psi)\equiv \psi$ we write
\begin{equation}
\label{gen-fn1} <\psi(1)>=\frac{\delta}{\delta {\xi}(1)}\ln Z_\xi~.
\end{equation}
with
\begin{equation}
\label{msr-ac2} Z_\xi = \int D\psi \int D\hat{\psi} \exp \left
[-{\cal A}_\xi [\psi,\hat{\psi}] \right ]~~.
\end{equation}
We have defined the generating functional $Z_\xi$ by including a
linear current term in the corresponding action ${\cal A}_\xi$
functional
\begin{equation}
\label{msr-ac6} {\cal A}_\xi [\psi,\hat{\psi}] = {\cal
A}[\psi,\hat{\psi}]- \int d1 \xi(1)\psi(1)~.
\end{equation}
The multi-point correlation functions of the variables $\psi$'s are
obtained from the generating functional,
\begin{equation}
\label{gen-fn} <\psi(1)....\psi(m)>=\frac{1}{Z_\xi}
\frac{\delta}{\delta \xi(1)} ..\frac{\delta}{\delta \xi(m)}Z_\xi
{\Big |}_{\xi=0}~.
\end{equation}
From the expression for the MSR action (\ref{msr-ac4}) and the
equation of motion (\ref{glgvn})  it follows that the linear part of
the dynamics produces a MSR action functional quadratic (Gaussian)
in the fields. Using the explicit forms of the equations of motion
(\ref{cont-eqn})-(\ref{conc-eqn}) we obtain the corresponding MSR
functional in the form
\begin{eqnarray}
\label{Aaction-MSR} {\cal A} &=& \int{dt}\int{d{\bf x}} \Big \{
\sum_{i,j} \hat{\rm g}_i\beta^{-1}L^0_{ij}\hat{\rm g}_j +
\hat{c}\beta^{-1}\gamma_{cc}\hat{c} + i\hat{\rho} \Big [
\frac{\partial\rho}{\partial t}+{\bf \nabla}
\cdot {\bf g} \Big ] \nonumber \\
&+& i\sum_{i} \hat{\rm g}_i \Big [ \frac{\partial {\rm
g}_i}{\partial t} + \rho \nabla_i \{ \chi^{-1}_{\rho\rho}\delta\rho
+\chi^{-1}_{\rho c}\delta c \}+c \nabla_i
\{ \chi^{-1}_{cc}\delta c+\chi^{-1}_{\rho c}\delta\rho \} \nonumber \\
&+&  \sum_j \nabla_j \{ {\rm g}_i \mathrm{v}_j \} + \sum_j L^0_{ij}
\mathrm{v}_j \Big ] + i\hat{c}\Big [ \frac{\partial c}{\partial t} +
\nabla_i\{cv_{i}\} +\gamma_0\nabla^2 \{ \chi^{-1}_{\rho c}\delta\rho
+\chi^{-1}_{cc}\delta c \} \Big ]
\Big \}~~. \nonumber \\
\end{eqnarray}
\subsubsection{Invariance of the MSR action}

Here we demonstrate that the MSR action functional given by Eq.
(\ref{Aaction-MSR}) is invariant under the transformation
(\ref{ttr-r1}) for the set of slow modes $\{\rho,c,{\bf g}\}$.
Changing $t$ to $-t$ in Eq. (\ref{Aaction-MSR}) and applying the
time transformation rules  for the above set and their corresponding
hatted counterparts, the action reduces to,
\begin{eqnarray}
\lbl{action-alg1}
&&{\cal A}[\psi(-t),\hat{\psi}(-t)] = \int d{\bf r} \int dt
\Big[\left\lbrace \hat{c}-i\beta\frac{\delta F}{\delta c}\right\rbrace
\beta^{-1}\gamma_{cc}\left\lbrace\hat{c}-i\beta\frac{\delta F}{\delta c}
\right\rbrace \\
&+& \left\lbrace-\hat{\rm g}_{i} + i\beta\frac{\delta F}{\delta {\rm
g}_{i}}\right\rbrace \beta^{-1}L_{ij}\left\lbrace-\hat{\rm g}_{i}+
i\beta\frac{\delta F}{\delta {\rm g}_{i}}\right\rbrace +
i\left\lbrace\hat{\rho}-i\beta\frac{\delta
F}{\delta\rho}\right\rbrace
\left\lbrace-\frac{\partial\rho}{\partial t}-Q_{\rho{{\rm
g}_i}}\frac{\delta F}{\delta{{\rm g}_i}}
\right\rbrace \nonumber \\
&+& i\left\lbrace \hat{c}-i\beta\frac{\delta F}{\delta c}
\right\rbrace\left\lbrace-\frac{\partial c}{\partial t}
-Q_{c{{\rm g}_i}}\frac{\delta F}{\delta {\rm g}_{i}}
+\gamma_{cc}\frac{\delta F}{\delta{c}}\right\rbrace\nonumber\\
&+& i\left\lbrace-\hat{\rm g}_{i}+i\beta\frac{\delta F}{\delta {\rm g}_{i}}
\right\rbrace\left\lbrace \frac{\partial {\rm g}_{i}}{\partial t}+
Q_{{\rm g}_i{{\rm g}_j}}\frac{\delta F}{\delta {\rm g}_{j}}
+ Q_{{\rm g}_i{\rho}}\frac{\delta F}{\delta{\rho}}
+Q_{{\rm g}_i{c}}\frac{\delta F}{\delta{c}}+ L_{ij}\frac{\delta F}{\delta {\rm g}_{j}}
\right\rbrace  \nonumber   \\
&=& {\cal A} [\psi,\hat{\psi}]+ i \beta \int_{-\infty}^{\infty} d1
\left\lbrace \frac{\delta F}{\delta
\psi_\alpha}Q_{\alpha\delta}\frac{\delta F}{\delta \psi_\delta}
\right\rbrace + \beta (F_{\infty}- F_{-\infty})~~. \nonumber
\end{eqnarray}
The second term inside the curly brackets on the RHS vanishes since
the dummy indices $\alpha$ and $\delta$ is summed over and the
Poisson bracket $Q_{\alpha\delta}$ is odd under the interchange of
the indices. The last term involving the $F$'s also vanish due to
equilibrium. Hence the MSR action is invariant under the time
reversal transformations defined by Eq. (\ref{ttr-r1}).

\subsection{Renormalized correlation functions}
\label{appendix3}

The construction of the field theoretical model fixes some basic
characteristics of the structure of the correlation and the response
functions. Let us first consider an important characteristic of the
Green's function matrix $G$ and $G_0$ in the MSR theory. For the
cases in which both indices in the matrix Eq. (\ref{full-Deqn})
correspond to the un-hatted fields, the following holds
\begin{itemize}

\item[(a)] $[{\bf G_0}^{-1}]_{\alpha\beta}=0$ which follows from the
action (\ref{Aaction-MSR}) obtained in the MSR field theory.

\item[(b)] ${\bf \Sigma}_{\alpha\beta} =0 $ which follows from causal nature of
the response functions in MSR field theory.

\end{itemize}
From the Schwinger-Dyson equation (\ref{full-Deqn}) we obtain that
the elements of the ${\bf G}^{-1}$ matrix corresponding to the
un-hatted fields, ${\left [{\bf G}^{-1}\right ]}_{\alpha\beta}=0$.
Inverting the matrix ${\bf G}^{-1}$ which has the above structure,
we obtain for the correlation functions of the physical, un-hatted
field variables,
\begin{equation}
 \label{Aeq:84} G_{\alpha\beta}=-\sum_{\mu\nu
} G_{\alpha\hat{\mu}}{\cal C}_{\hat{\mu}\hat{\nu}}
G_{\hat{\nu}\beta}~,
\end{equation}
where Greek letter subscripts take values $\rho ,c$ and the
longitudinal components of the vector field  ${\bf g}$. The matrix
${\cal C}_{\hat{\mu}\hat{\nu}}$ is given by,
\begin{equation}\label{Ac-matrix}
{\cal C}_{\hat{\mu}\hat{\nu}}= {[{\cal C}_0]}_{\hat{\mu}\hat{\nu}}
-\Sigma_{\hat{\mu}\hat{\nu}}~~,
\end{equation}
and is listed in Table \ref{table2}. The double-hatted self-energies
$\Sigma_{\hat{\mu}\hat{\nu}}$ as well as ${\cal
C}_{\hat{\mu}\hat{\nu}}$ both vanish if either index corresponds to
$\hat{\rho}$, since there is no noise or nonlinearity in the
continuity equation (\ref{cont-eqn}).

The response part of $G_{\hat{\alpha}\beta}^{-1}$ is obtained using
Eq. (\ref{full-Deqn}) in terms of the corresponding elements of the
matrix of $[{\cal B}_0-\Sigma]$. The elements of
$G_{\hat{\alpha}\beta}^{-1}$ matrix are listed in Table
\ref{table3}. The renormalized response function
$G_{\alpha\hat{\varphi}}$ is obtained in the form
\be \label{Aresp-matx} G_{\alpha\hat{\varphi}}=
\frac{N_{\alpha\hat{\varphi}}}{{\cal D}}~, \ee
where the elements of matrix $N_{{\alpha}\hat{\varphi}}$ are given
in Table \ref{table4}. The denominator ${\cal D}$ in the RHS of Eq.
(\ref{resp-matx}) is obtained as,
\begin{equation}
\label{denom-resp} {\cal D}({\bf q},\omega)= (\omega+iq^2\nu) \left
[\omega^2-q^{2}c^{2} +i{\omega}L q^2\right ]
+iq^{4}\upsilon^{2}{\nu_1}~~.
\end{equation}
The various renormalized transport coefficients  appearing on the
RHS of Eq. (\ref{denom-resp}) are expressed in terms of the
corresponding response self-energies,
\bea \label{renorm-visc}
Lq^2 &=& L^0q^2+i\Sigma_{\hat{\rm g}{\rm g}}~, \\
\lbl{renorm-nu}
\nu q^2 &=& \gamma_0 \chi^{-1}_{c{c}}q^2+i\Sigma_{\hat{c}c}~, \\
\label{renorm-nup} \nu'q^2 &=& \gamma_0 \chi^{-1}_{\rho c}q^2
+i\Sigma_{\hat{c}\rho}~. \eea
The quantity $\nu_1$ in the last term on the RHS of the definition
(\ref{denom-resp}) is obtained as $\nu_1=\nu'+
\omega\gamma_{\hat{c}{\rm g}}$ in terms of the leading order nonzero
contributions to the self-energy $\Sigma_{\hat{c}{\rm g}} =
-iq^3\gamma_{\hat{c}{\rm g}}$ in the small $q$ limit. $c^2$ and
$\upsilon^2$ respectively represent the renormalized expressions for
the sound speeds $c_0^2=\rho_0\chi_{\rho\rho}^{-1}$ and
$\upsilon_0^2=\rho_0\chi_{\rho{c}}^{-1}$ obtained in terms of the
self-energies
\bea \label{renorm-s1}
c^{2} = c_{0}^{2}+q^{-1}\Sigma_{\hat{\rm g}\rho}~, \\
\lbl{renorm-s2} \upsilon^{2}=\upsilon_{0}^{2}+
q^{-1}\Sigma_{\hat{\rm g}c}~. \eea

\subsection{Analysis of the FDRs}
\label{appendix4}

We begin with the Fluctuation dissipation relation (\ref{ttr-gt})
corresponding to $\psi=g$,
\be G_{{\rm g}_{i}\varphi}({\bf q},\omega)= -2\beta^{-1}\rho_{0}Im
G_{\hat{\rm g}_{i}\varphi}({\bf q},\omega). \ee
Using the definitions (\ref{resp-matx} ) and (\ref{eq:84}) for the
correlation and response functions respectively in the above
fluctuation-dissipation relation we obtain the result
\begin{eqnarray}
& &\sum_{\hat{\alpha},\hat{\gamma}} G_{{\rm g}_{i}\hat{\alpha}}{\cal
C}_{\hat{\alpha}\hat{\gamma}}G_{\hat{\gamma}\varphi}=
2\beta^{-1}\rho_{0}\mathrm{Im} G_{\hat{\rm g}_{i}\varphi}~~,
\nonumber\\
& &\sum_{\hat{\alpha}}G_{{\rm g}_{i}\hat{\alpha}}{\cal
C}_{\hat{\alpha}\hat{\gamma}}=-i\beta^{-1}\rho_{0} \sum_{\beta}\Big
[ G_{\hat{\rm g}_{i}\varphi}-G_{\hat{\rm g}_{i}\varphi}^{*} \big ]
G_{\varphi\hat{\gamma}}^{-1},
\nonumber\\
& &\label{fdtnc} \sum_{\hat{\alpha}}N_{{\rm g}_{i}\hat{\alpha}}{\cal
C}_{\hat{\alpha}\hat{\gamma}} =-i\beta^{-1}\rho_{0}({\cal
D}\delta_{\hat {\rm g}_{i}\hat{\gamma}}+\sum_{\beta}N^*_{\hat{\rm
g}_{j}\varphi}G^{-1}_{\varphi\hat{\gamma}}) ~~~.
\end{eqnarray}
On substituting $\hat{\gamma}=\hat{c}$ and $\hat{\rm g}$ in Eq.
(\ref{fdtnc}), we obtain respectively the following equations:
\bea N_{{\rm g}\hat{c}}{\cal C}_{\hat{c}\hat{c}}+N_{{\rm g}\hat{\rm
g}}{\cal C}_{\hat{\rm g}\hat{c}} &=&
-i\beta^{-1}\rho_{0}(N_{\hat{\rm
g}\rho}^{*}G_{\rho\hat{c}}^{-1}+N_{\hat{\rm
g}c}^{*}G_{c\hat{c}}^{-1}+N_{\hat{\rm g}{\rm g}}^{*}
G_{{\rm g}\hat{c}}^{-1})~~,\\
N_{{\rm g}\hat{c}}{\cal C}_{\hat{c}\hat{\rm g}}+N_{{\rm g}\hat{\rm
g}}{\cal C}_{\hat{\rm g}\hat{\rm g}} &=& -i\beta^{-1}\rho_{0}( {\cal
D}+N_{\hat{\rm g}\rho}^{*}G_{\rho\hat{\rm g}}^{-1} +N_{\hat{\rm
g}c}^{*}G_{c\hat{\rm g}}^{-1}+N_{\hat{\rm g}{\rm g}} ^{*}G_{{\rm
g}\hat{\rm g}}^{-1}). \eea
Equating the real and imaginary parts from both sides and using the
fact that the elements ${\cal C}_{\hat{\rm g}\hat{\rm g}}$ and
${\cal C}_{\hat{c}\hat{c}}$ are real while ${\cal C}_{\hat{\rm
g}\hat{c}}={\cal C}_{\hat{c}\hat{\rm g}}^{*}$ are not, we obtain a
set of relations between the correlation and response self-energies.
For the self-energy elements ${\cal C}_{\hat{\rm g}\hat{\rm g}}$ and
${\cal C}_{\hat{c}\hat{c}}$ we respectively obtain the following
results
\begin{eqnarray}
\lbl{fnl-r1}
{\cal C}_{\hat{\rm g}\hat{\rm g}}+\frac{{|N_{{\rm
g}\hat{c}}|}^2}{{\cal M}}{\cal C}_{\hat{c}\hat{\rm g}}^{\prime} &=&
2\beta^{-1}\rho_0\Big [ \frac{N_{{\rm g}\hat{c}}^{\prime}}{{\cal M}}
\Big \{ N_{\hat{\rm g}{\rm g}}^{\prime}G_{g\hat{\rm
g}}^{-1\prime\prime}+ N_{\hat{\rm g}c}^{\prime}G_{c\hat{\rm
g}}^{-1\prime\prime}+ N_{\hat{\rm g}\rho}^{\prime}G_{\rho\hat{\rm
g}}^{-1\prime\prime}\Big \}  \nonumber \\
&+& \frac{N_{{\rm g}\hat{c}}^{\prime\prime}}{{\cal M}} \Big \{
N_{\hat{\rm g}{\rm g}}^{\prime\prime} G_{{\rm g}\hat{\rm
g}}^{-1\prime\prime}+N_{\hat{\rm g}c}^{\prime\prime} G_{c\hat{\rm
g}}^{-1\prime\prime}+N_{\hat{\rm
g}\rho}^{\prime\prime}G_{\rho\hat{\rm g}}^{-1\prime\prime} \Big \}
\Big ]~~, \\
\label{fnl-r2} {\cal C}_{\hat{c}\hat{c}}+\frac{{|N_{{\rm g}\hat{\rm
g}}|}^2}{{\cal M}} {\cal C}_{\hat{c}\hat{\rm g}}^{\prime} &=&
2\beta^{-1}\rho_0\Big [\frac{N_{{\rm g}\hat{\rm g}}^{\prime}}{{\cal
M}} \{ N_{\hat{\rm g}\rho}^{\prime}G_{\rho\hat{c}}^{-1\prime\prime}-
N_{\hat{\rm
g}\rho}^{\prime\prime}G_{\rho\hat{c}}^{-1\prime}+N_{\hat{\rm g}{\rm
g}}^{\prime}G_{{\rm g}\hat{c}}^{-1\prime\prime} -N_{\hat{\rm g}{\rm
g}}^{\prime\prime}G_{{\rm g}\hat{c}}^{-1\prime}\} \Big ]~,
\end{eqnarray}
with ${\cal M}= N_{{\rm g}\hat{\rm g}}^{\prime}N_{{\rm
g}\hat{c}}^{\prime} +N_{{\rm g}\hat{\rm g}}^{\prime\prime}N_{{\rm
g}\hat{c}}^{\prime\prime}$~.

Next we consider the FDR (\ref{fdt-q2}) to obtain another set of
relations between the correlation  and response self-energies.
Evaluating the functional derivatives $\zeta_c$ for a Gaussian free
energy we obtain the result,
\be \label{fdt-a21} \chi_{\rho c}^{-1}G_{\rho\varphi}({\bf
q},\omega)+\chi_{cc}^{-1}G_{c\varphi}({\bf
q},\omega)=-2\beta^{-1}\mathrm{Im} G_{\hat{c}\varphi}({\bf
q},\omega)~. \ee
Following the same procedures as in the case of the FDR  the Eq.
(\ref{fdt-a21}) reduces to the form
\be \label{fdt-a22} \sum_{\hat{\alpha}}\Big [ \chi_{\rho
c}^{-1}N_{\rho\hat{\alpha}}{\cal
C}_{\hat{\alpha}\hat{\gamma}}+\chi_{cc}^{-1} N_{c\hat{\delta}}{\cal
C}_{\hat{\delta}\hat{\gamma}} \Big ] =-i\beta^{-1} \{ {\cal
D}\delta_{\hat{c}\hat{\gamma}} + \sum_{\beta}
N_{\hat{c}\varphi}^*G_{\varphi\hat{\gamma}}^{-1} \}~ . \ee
Setting  $\hat{\gamma}=\hat{c}$ in Eq. (\ref{fdt-a22}) we obtain
\begin{eqnarray}
\label{fdt-a23a} &&\chi_{\rho c}^{-1}(N_{\rho \hat{c}}{\cal
C}_{\hat{c}\hat{c}} +N_{\rho\hat{\rm g}}{\cal C}_{\hat{\rm
g}\hat{c}})+ \chi_{cc}^{-1}(N_{c\hat{c}}{\cal C}_{\hat{c}\hat{c}}
+N_{c\hat{\rm g}}{\cal C}_{\hat{\rm g}\hat{c}}) \nonumber \\
&=&-i\beta^{-1} \Big [ {\cal D}+
N^*_{\hat{c}\rho}G_{\rho\hat{c}}^{-1}+N^*_{\hat{c}c}G_{c\hat{c}}^{-1}
+N^*_{\hat{c}{\rm g}}G_{{\rm g}\hat{c}}^{-1}) \Big ]~.
\label{fdt-a23}
\end{eqnarray}
Next, substituting $\hat{\gamma}=\hat{\rm g}$ in Eq.
(\ref{fdt-a22}) we obtain
\begin{eqnarray}
&&\label{fdt-a24}\chi_{\rho c}^{-1}(N_{\rho \hat{c}}{\cal
C}_{\hat{c}\hat{\rm g}}+N_{\rho\hat{\rm g}}{\cal C}_{\hat{\rm
g}\hat{\rm g}})+\chi_{cc}^{-1}(N_{c\hat{c}}{\cal C}_{\hat{c}\hat{\rm
g}}+N_{c\hat{\rm g}}{\cal C}_{\hat{\rm g}\hat{\rm g}}) \nonumber
\\&=& -i\beta^{-1} \Big [ N_{\hat{c}\rho}G_{\rho\hat{\rm
g}}^{-1}+N_{\hat{c}c}G_{c\hat{\rm g}}^{-1} +N_{\hat{c}{\rm
g}}G_{g\hat{\rm g}}^{-1}) \Big ]~.
\end{eqnarray}
Comparing real and imaginary parts of Eqs. (\ref{fdt-a23}) and
(\ref{fdt-a24}), we obtain the following results respectively for
${\cal C}_{\hat{c}\hat{c}}$ and ${\cal C}_{\hat{g}\hat{g}}$
\begin{eqnarray}
{\cal C}_{\hat{c}\hat{c}}+\frac{{|{\cal J}|}^2}{{\cal
Q}} {\cal C}_{\hat{c}\hat{\rm g}}^{\prime} &=&2\beta^{-1}\Bigg [
\frac{{\cal J}^\prime}{{\cal Q}} \Big \{
N_{\hat{c}\rho}^{\prime}G_{\rho\hat{c}}^{-1\prime\prime}
+N_{\hat{c}c}^{\prime}G_{c\hat{c}}^{-1\prime\prime}
+N_{\hat{c}{\rm g}}^{\prime}G_{{\rm g}\hat{c}}^{-1\prime\prime} \Big \}
\nonumber \\
\label{fnl-r3} &+& \frac{{\cal J}^{\prime\prime}}{{\cal Q}} \Big \{
N_{\hat{c}\rho}^{\prime\prime}G_{\rho\hat{c}}^{-1\prime\prime}
+N_{\hat{c}c}^{\prime\prime}G_{c\hat{c}}^{-1\prime\prime}
+N_{\hat{c}{\rm g}}^{\prime\prime}G_{{\rm
g}\hat{c}}^{-1\prime\prime} \Big \}\Bigg ]  , \\
\label{fnl-r4} {\cal C}_{\hat{\rm g}\hat{\rm g}}+\frac{{|{\cal
K}|}^2}{{\cal Q}}{\cal C}_{\hat{c}\hat{\rm
g}}^{\prime}&=&2\beta^{-1}\Bigg [ \frac{{\cal K}^\prime}{{\cal Q}}
\Big \{ N_{\hat{c}\rho}^{\prime}G_{\rho\hat{\rm g}}^{-1\prime\prime}
-N_{\hat{c}\rho}^{\prime\prime}G_{\rho\hat{\rm g}}^{-1\prime}
+N_{\hat{c}{\rm g}}^{\prime}G_{{\rm g}\hat{\rm g}}^{-1\prime\prime}
-N_{\hat{c}{\rm g}}^{\prime\prime}G_{{\rm g}\hat{\rm g}}^{-1\prime}
\Big \} \Bigg ]~~~.
\end{eqnarray}
In the above equations we have defined the quantities ${\cal
J},{\cal K},$ and ${\cal Q}$ in terms of the matrix elements of
$N_{\alpha\hat{\beta}}$ and $\chi^{-1}_{\alpha\beta}$ as follows:
\bea \label{rel-def1} {\cal J} &=& \chi_{\rho c}^{-1}N_{\rho
\hat{\rm g}}
+\chi_{cc}^{-1}N_{c\hat{\rm g}} ~,\\
\label{rel-def2} {\cal K} &=& \chi_{\rho c}^{-1}N_{\rho \hat{c}}
+\chi_{cc}^{-1}N_{c\hat{c}}~, \\
\label{rel-def3} {\cal Q}&=&{\cal J}^{\prime}{\cal K}^{\prime}
+{\cal J}^{\prime\prime}{\cal K}^{\prime\prime}~~. \eea
%
In general, for finite wave number ($q$) and frequency ($\omega$)
the above FDRs (\ref{fnl-r1}), (\ref{fnl-r2}), (\ref{fnl-r3}), and
(\ref{fnl-r4}) between the real and imaginary parts of the
correlation and response self-energies are complicated and difficult
to resolve. Here we analyze their implications in the hydrodynamic
limit of small $q$ and $\omega$ by writing the dependence of the
various self-energy elements to leading order in the wave number $q$
using simple symmetry arguments.

\subsection{Self-energy relations}

We begin by considering the correlation self-energy matrix element
$\Sigma_{\hat{\rm g}_i\hat{\rm g}_j}({\bf q},\omega)$. From the MSR
action functional for the two component mixture given by Eq.
(\ref{Aaction-MSR}), it follows that the cubic vertices with a
$\hat{{\rm g}_{i}}(q)$ leg each, contribute an explicit $q_{i}$
factor. The self-energy $\Sigma_{\hat{\rm g}_{i}\hat{\rm
g}_{j}}(q,\omega)$ has two vertices with external legs of $\hat{\rm
g}_{i}(q)$ and $\hat{\rm g}_{j}(q)$. Hence this self-energy involves
an explicit factor of $q_iq_j$ {\em i.e.}, $\Sigma_{\hat{\rm
g}_{i}\hat{\rm g}_{j}}\sim -q_iq_j \gamma_{\hat{\rm g}\hat{\rm g}}
\equiv -q^2\gamma_{\hat{\rm g}_{i}\hat{\rm g}_{j}}$. Let us now
consider the self-energy element
$\Sigma_{\hat{c}\hat{c}}(q,\omega)$. The cubic vertex with a
$\hat{c}(q)$ leg contributes a $q$ factor. Hence, using similar
arguments as above, the self-energy
$\Sigma_{\hat{c}\hat{c}}(q,\omega)$ also involves an explicit factor
of $q^2$. However in this case the vector indices must be contracted
to produce a scalar form. Therefore we obtain,
$\Sigma_{\hat{c}\hat{c}}\sim -q^2 \gamma_{\hat{c}\hat{c}}$. We also
verify these results explicitly at the one loop order by considering
the corresponding diagrams for the $\Sigma_{\hat{\rm g}_{i}\hat{\rm
g}_{j}}$ and $\Sigma_{\hat{c}\hat{c}}$.

Next, we consider the response self-energy $\Sigma_{\hat{\rm
g}_{i}{\rm g}_{j}}({\bf q},\omega)$ which contains only one external
$\hat{{\rm g}_{i}}$ contributing a factor $q_{i}$. The other leg of
this self-energy  involves the vector field ${\rm g}_j$ and hence
the $O(q)$ level contribution must have the  external $q_i$ factor
(due to the hatted field $\hat{\rm g}_i (q)$) multiplied to an
explicit internal $k_j$ wave-vector.  The internal wave-vector is
integrated out. At this $O(q)$, the external $q$  is set equal to
zero in the integral for the diagrammatic contribution. However this
integral vanishes for being  odd in $k$, due to the ${\bf
k}\rightarrow{-{\bf k}}$ symmetry. Hence this response self-energy
is at least of the $O(q^2)$ and writing this out explicitly we
obtain $\Sigma_{\hat{\rm g}_{i}\hat{\rm g}_{j}}\sim -iq^2
\gamma_{\hat{\rm g}_{i}{\rm g}_j}$. Using exactly similar arguments
for the self-energy $\Sigma_{\hat{c}c}({\bf q},\omega)$, we note
that since $\hat{c}$ is a scalar field the $O(q)$ contribution must
have the external $q$ factor contracted to an explicit $k$ internal
wave-vector which is integrated out. Again since at this order the
external $q$ is  set equal to zero, the integral vanishes being odd
in $k$. Hence we define taking into account this factor
$\Sigma_{\hat{c}c}\sim -iq^2 \gamma_{\hat{c}c}$. We also verify
these behaviors explicitly at the one loop order by considering the
diagrams for the $\Sigma_{\hat{\rm g}_{i}{\rm g}_{j}}$ and
$\Sigma_{\hat{c}{c}}$.

Next the self-energy $\Sigma_{\hat{\rm g}_i\rho}({\bf q},\omega)$ is
considered. Due to the the external leg $\hat{\rm g}_i$ in a vertex
an explicit factor of $q_i$ appear. To consider the $O(q)$
contribution of this self-energy  we therefore set $q=0$ in all the
internal integrations and thus this contribution vanishes. We can
establish this result at the one loop order by considering the
detailed nature of the vertices which contribute to this
self-energy. These are of the following two types: a) those which
has one $\rho$ leg. Such vertices in the present model has {\em one
other leg} with a vector index ${\rm g}_i$ or $\hat{\rm g}_i$ (For
example, the vertices $V_{\hat{\rm g}_i\rho\rho}$, $V_{\hat{\rm
g}_i\rho{c}}$, and $V_{\hat{c}{{\rm g}_i}\rho}$ ). b) those which
has one $\hat{\rm g}_i$ leg. Such vertices in the present model has
{\em two other legs} each with a vector index ${\rm g}_j$ etc.(For
example, the vertices $V_{\hat{\rm g}_i{{\rm g}_j}{{\rm g}_i}}$), or
without {\em any other leg} having a vector index (for example
$V_{\hat{\rm g}_i\rho{c}}$, $V_{\hat{\rm g}_i\rho\rho}$, and
$V_{\hat{\rm g}_i{c}{c}}$ ). As a result of this, the internal
integration involved in the one loop diagrammatic contribution
involves either one or three powers of the internal vector ${\bf
k}$. For the isotropic liquid such an integral must vanish, since
using the ${\bf k} \rightarrow -{\bf k}$ symmetry the integrand is
an odd function. Thus the $O(q)$ contribution to $\Sigma_{\hat{\rm
g}_i\rho}({\bf q},\omega)$ is taken to be zero. The next order
surviving contribution must therefore be $O(q_iq^2)$. Hence we write
this self-energy as
\be
\Sigma_{\hat{\rm g}_i\rho}({\bf q},\omega)=
-iq_iq^2 \gamma_{\hat{\rm g}\rho}({\bf q},\omega)
\equiv -iq^3 \gamma_{\hat{\rm g}_i{\rho}}({\bf q},\omega).
\ee
In a similar way we can show that $\Sigma_{\hat{\rm g}_i{c}}({\bf
q},\omega) \sim -iq^3 \gamma_{\hat{\rm g}_{i}{c}}({\bf q},\omega)$.
Finally, the  self-energy $\Sigma_{\hat{c}\hat{\rm g}_{i}}$ has a
factor $q_i$ due to the external leg $\hat{\rm g}_i$ and a factor of
$q_j$ due to the leg $\hat{c}$. The latter must be contracted with
an internal $k_j$ vector. Hence the $O(q_iq_j)$
contribution involves an integral which is odd in 
. The latter vanishes making the lowest order contribution to the
self-energy being of $O(q_iq^2)$. Therefore we write
$\Sigma_{\hat{c}\hat{\rm g}_i}({\bf q},\omega)=-iq_iq^2
\gamma_{\hat{c}\hat{\rm g}}({\bf q},\omega) \equiv -iq^3
\gamma_{\hat{c}\hat{\rm g}_i}({\bf q},\omega)$. The leading order
contributions to these self-energies are listed in Table
\ref{table5}. Substituting the relevant elements of matrix
$G^{-1}_{\alpha\hat{\beta}}$ and matrix $N_{\hat{\alpha}\beta}$ in
Eqs. (\ref{fnl-r1}) and (\ref{fnl-r2}), we obtain by comparing
leading order terms the following relations between the correlation
and response type self-energies
\bea
\lbl{hydsrel-e1}
\gamma_{\hat{c}\hat{c}} &=&
2\beta^{-1}\frac{\gamma_{\hat{c}\rho}^{\prime}}{\chi_{\rho c}^{-1}}~, \\
\lbl{hydsrel-e2}
\gamma_{\hat{\rm g}\hat{\rm g}} &=& 2\beta^{-1}\rho_{0}
\gamma_{\hat{\rm g}\rm g}^{\prime}~,
\eea
and the comparison of next higher order terms from these two
equations respectively obtain the self-energy relations
\bea
\lbl{hydsrel-e3}
\gamma_{\hat{c}\hat{\rm g}} &=& 2\beta^{-1}\rho_{0}
\gamma_{\hat{c}\rm g}^{\prime}~,\\
\lbl{hydsrel-e4}
\gamma_{\hat{c}\hat{\rm g}} &=&
2\beta^{-1}\frac{\gamma_{\hat{\rm g}\rho}^{\prime}}{\chi_{\rho c}^{-1}}~.
\eea
In an exactly similar way, using  elements of matrix
$G^{-1}_{\alpha\hat{\beta}}$ and matrix $N_{\hat{\alpha}\beta}$ in
Eqs. (\ref{fnl-r3}) and (\ref{fnl-r4}) and by comparing leading
order terms the following relations between the correlation  and
response type self-energies, we obtain \bea \lbl{hydsrel-e1r}
\gamma_{\hat{c}\hat{c}}&=& 2\beta^{-1}
\frac{\gamma_{\hat{c}c}^{\prime}}{\chi_{cc}^{-1}}~,\\
\lbl{hydsrel-e2r}
\gamma_{\hat{\rm g}\hat{\rm g}}&=&2\beta^{-1}\rho_{0}
\gamma_{\hat{\rm g}\rm g}^{\prime}~.
\eea
Note that we have now reached the relation between $\gamma_{\hat{\rm
g}\hat{\rm g}}$ with the corresponding response self-energy
$\gamma_{\hat{\rm g}\rm g}$ ( Eqs. (\ref{hydsrel-e2}) and
(\ref{hydsrel-e2r})) and either of these self-energies can be used
to renormalize the longitudinal viscosity. On the other hand the
Eqs. (\ref{hydsrel-e1}) and (\ref{hydsrel-e1r}) link two response
self-energies $\gamma_{\hat{c}\rho}$ and $\gamma_{\hat{c}c}$ to a
single self-energy $\gamma_{\hat{c}\hat{c}}$ as
\be
\lbl{com-sreln}
\gamma_{\hat{c}\hat{c}} =
2\beta^{-1}\frac{\gamma_{\hat{c}\rho}^{\prime}}{\chi_{\rho c}^{-1}}=
2\beta^{-1}\frac{\gamma_{\hat{c}c}^{\prime}}{\chi_{cc}^{-1}}~~.
\ee
This proves an important relation by which two different transport
coefficients are renormalized in terms of the same self-energy
$\gamma_{\hat{c}\hat{c}}$. Finally, comparing the next order terms
from Eq. (\ref{fnl-r4}) and making use of the Eq.
(\ref{hydsrel-e4}) linking $\gamma_{\hat{\rm g}\rho}^{\prime}$ with
$\gamma_{\hat{c}\hat{\rm g}}$, we obtain the result
\begin{equation}
\gamma_{\hat{c}\hat{\rm g}}=2\beta^{-1}\frac{\gamma_{\hat{\rm
g}c}^{\prime}}{\chi_{cc}^{-1}}~.
\end{equation}

\newpage

\begin{table}[!htbp]
\begin{center}
\begin{tabular}{|c|ccc|}\hline
~~&~~ ${\rho}$~~ ~~&~~~~ ${c}$~~ ~~&~~~~ ${\rm g}$ \\\hline
~~&~~&~&~~\\
$\hat{\rho}$~~&~~~~$\omega$~~~~&~~~~$0$~~~~&~~~~$-q$~~\\
$~$~~~~&~~~~$~$~~~~&~~~~$~$~~~~&~~~~$~$~~\\
$\hat{c}$~~&~~~~$iq^{2}\nu_0^\prime$~~~~&~~~~$\omega+iq^2\nu_0$ ~~&~~$0$~~\\
~~~~&~~~~~~&~~~~~~&~~~~~\\
$\hat{\rm g}$~~&~~~~$-qc_0^{2}$~~~~&~~~~$-q\upsilon_0^{2}$
~~~~&~~~~$\omega+iq^2L^0~~$\\
~~&~~&~&~~\\\hline
\end{tabular}
\caption{Elements of matrix ${[G_0^{-1}]}_{\hat{\alpha}\beta}$
defined in terms of the matrix ${\cal B}_0$.
$\nu_0^\prime=\chi_{\rho{c}}^{-1}\gamma_0$, and $\nu_0={\chi}_{c
c}^{-1}\gamma_0$.} \label{table1}
\end{center}
\end{table}


\begin{table}[!htbp]
\begin{center}
\begin{tabular}{|c|ccc|}\hline
~~&~~ $~~~~\hat{\rho}$~~~~ ~~&~~~~ $\hat{c}$~~ ~&~~ $\hat{\rm g}$
\\\hline $~$~~~~&~~~~$~$~~~~&~~~~$~$~~~~&~~~~$~$~~\\
$~~~~~\hat{\rho}$~~~~~~&~~~~$0$~~~&~~$0$~~~&~~$0$~~\\
$~$~~~&~~$~$~~~&~~$~$~~~&~~$~$~~\\
$~~~~\hat{c}$~~~~~~&~~~~$0$~~~~&~~~~~~$2\beta^{-1}q^2\gamma_{0}
-\Sigma_{\hat{c}\hat{c}}$
~~~~&~~~~$-\Sigma_{\hat{c}\hat{{\rm g}}}$~~\\
$~$~~~~&~~~~$~$~~~~&~~~~$~$~~~~&~~~~$~$~~\\
$\hat{\rm g}$~~&~~~~$0$~~~~&~~~~$-\Sigma_{\hat{{\rm g}}\hat{c}}$
~~~~&~~~~$2\beta^{-1}q^2L_{0}-\Sigma_{\hat{{\rm g}}\hat{{\rm g}}}$\\
~~&~~&~&~~\\\hline
\end{tabular}
\caption{Elements of matrix ${\cal C}_{\hat{\alpha}\hat{\beta}}$.}
\label{table2}
\end{center}
\end{table}


\begin{table}[!htbp]
\begin{center}
\begin{tabular}
{|c|ccc|}\hline ~~&~~ ${\rho}$~~ ~~&~~~~ ${c}$~~ ~~&~~~~ ${\rm g}$
\\\hline
~~&~~&~&~~\\
$\hat{\rho}$~~&~~~~$\omega$~~~~&~~~~$0$~~~~&~~~~$-q$~~\\
$~$~~~~&~~~~$~$~~~~&~~~~$~$~~~~&~~~~$~$~~\\
$\hat{c}$~~&~~~~$iq^{2}\nu^\prime$~~~~&~~~~$\omega+iq^2\nu
$ ~~~~&~~~~$-\Sigma_{\hat{c}{\rm g}}$~~\\
$~$~~~~&~~~~$~$~~~~&~~~~$~$~~~~&~~~~$~$~~\\
$\hat{\rm g}$~~&~~~~$-qc^{2}$~~~~&~~~~$-q\upsilon^{2}$
~~~~&~~~~$\omega+iq^2L~~$\\
~~&~~&~&~~\\\hline
\end{tabular}
\caption{Elements of matrix $G^{-1}_{\hat{\alpha}\beta}$ in terms of
the renormalized transport coefficients $L$, $\nu$, and $\nu^\prime$
respectively defined in Eqs. (\ref{renp-visc}) -(\ref{renp-mup}).
The symbols $\upsilon^2$, $c^2$ are explained in the text.}
\label{table3}
\end{center}
\end{table}


\begin{table}[!htbp]
 \begin{center}
\begin{tabular}{|c|ccc|}\hline
 ~~&~~ $\hat{\rho}$~~ ~~&~~~~ $\hat{c}$~~ ~~&~~~~ $\hat{\rm g}$ \\\hline
$~$~~~~&~~~~$~$~~~~&~~~~$~$~~~~&~~~~$~$~~\\
 $\rho$~~&~~~~$(\omega+iq^2L~~)(\omega+iq^2\nu ~)
$~~~~&~~~~$q^2\upsilon^{2}$~~~~&~~~~$q(\omega+iq^2\nu ~)$~~\\
$~$~~~~&~~~~$+iq^4\upsilon^2\gamma_{\hat{c}{\rm g}}$~~~~&~~~~$~$~~~~&~~~~$~$~~\\
$~$~~~~&~~~~$~$~~~~&~~~~$~$~~~~&~~~~$~$~~\\
 $c$~~&~~~~$-iq^{2}\{\nu^\prime (\omega+iq^2L)~~
$~~~~&~~~~$\omega (\omega+iq^2L)~~$
~~~~&~~~~$-iq^{3}(\nu^\prime+\omega\gamma_{\hat{c}{\rm g}})$~~\\
$~$~~~~&~~~~$+q^2c^{2}\gamma_{\hat{c}{\rm g}}\}$~~~~&~~~~$-q^{2}
c^{2}$~~~~&~~~~$~$~~\\
$~$~~~~&~~~~$~$~~~~&~~~~$~$~~~~&~~~~$~$~~\\
 $g$~~&~~~~$-iq^{3}\upsilon^{2}\nu^\prime
+qc^{2}(\omega+iq^2\nu ~)$~~~~&~~~~$\omega q\upsilon^{2}$
 ~~~~&~~~~$\omega (\omega+iq^2\nu ~)$\\
~~&~~&~&~~\\\hline
\end{tabular}
\caption{Elements of matrix $N_{\alpha\hat{\beta}}$ in terms of the
renormalized transport coefficients $L$, $\nu$, and $\nu^\prime$
respectively defined in Eqs. (\ref{renp-visc}) -(\ref{renp-mup}).
The symbols $\upsilon^2$, $c^2$ are explained in the text.
$\gamma_{\hat{c}{\rm g}}$ is the leading order contribution to the
corresponding self-energy $\Sigma_{\hat{c}{\rm g}} =
-iq^3\gamma_{\hat{c}{\rm g}}$.} \label{table4}
\end{center}
\end{table}


\begin{table}
\begin{center}
\begin{tabular}{|c|ccccc|}\hline
 ~~&~~$\rho$~~&~~$c$~~&~~${\rm g}_{j}$~~&~~$\hat{c}$ ~~&~~
$\hat{\rm g}_{j}$\\ \hline
~~&~~&~&~~&~&~~\\
$\hat{c}$ ~~&~~$-iq^2\gamma_{\hat{c}\rho}$~~&~~$-iq^2
\gamma_{\hat{c}c}$~~&~~$-iq^3\gamma_{\hat{c}{\rm g}_{j}}$ ~~&~~
$-q^2\gamma_{\hat{c}\hat{c}}$ ~~&~~ $-q^3
\gamma_{\hat{c}\hat{\rm g}_{j}}$ \\
~~&~~&~&~~&~&~~\\
$\hat{\rm g}_{i}$ ~~&~~$-iq^3\gamma_{\hat{\rm g}_{i}\rho}$~~
&~~$-iq^3\gamma_{\hat{\rm g}_{i}c}$ ~~&~~$-iq^2\gamma_{\hat{\rm
g}_{i}{\rm g}_{j}}$~~ &~~$-q^3\gamma_{\hat{\rm g}_{i}\hat{c}}$ ~~&~~
 $-q^2\gamma_{\hat{\rm g}_{i}\hat{\rm g}_{j}}$ \\
~~~&~~&~&~~&~&~~\\\hline
 \end{tabular}
\caption{$q$ dependence of self-energies
$\Sigma_{\hat{\alpha}\beta}$ and
$\Sigma_{\hat{\alpha}\hat{\beta}}$.} \label{table5}
\end{center}
\end{table}

\end{document}